\documentclass[aps,prd,twocolumn,superscriptaddress,nofootinbib,longbibliography]{revtex4-2}
\usepackage{amsmath,amssymb,amsthm}
\usepackage{mathtools}
\usepackage{physics}
\usepackage{bm}
\usepackage{tensor}
\usepackage{xspace}
\usepackage{empheq}
\usepackage{tabularx}
\usepackage{graphicx}
\usepackage{booktabs}
\usepackage{array}
\usepackage[colorlinks=true,allcolors=blue]{hyperref}
\usepackage{float}
\usepackage{comment}
\usepackage{booktabs}

\newcommand{\pRR}{{}^{*}\!R R}
\newcommand{\dual}[1]{{}^{*}\!#1}

\newcommand{\cell}[1]{\begin{tabular}[t]{@{}l@{}}#1\end{tabular}}

\begin{document}

\title{
Isospectrality breaking in the axial quasinormal modes of Reissner–Nordström black holes in dynamical Chern–Simons gravity
}

\author{Abhishek Rout}
\email{arout@email.sc.edu}

\affiliation{Department of Physics and Astronomy,\\ University of South Carolina, Columbia, South Carolina 29208, USA}

\date{\today}

\begin{abstract}
We study odd-parity perturbations of Reissner–Nordström black holes in dynamical Chern–Simons gravity. The dCS action is parity invariant because $\vartheta \pRR$ is the product of two pseudoscalars; nevertheless, the pseudoscalar perturbation has odd parity and enters only the axial sector at linear order, and so lifts the axial--polar isospectrality of general relativity. We derive the finite-charge couplings between the pseudoscalar and the axial gravitational and electromagnetic perturbations directly from the linearized field equations. Both directions are governed by the radial kernel $(Mr-Q^2)\Delta/r^8$, and at nonzero charge the pseudoscalar couples to $\Psi+\chi\Psi_e$, with $\chi=2Q/[r\sqrt{(l-1)(l+2)}]$. After diagonalizing the Einstein--Maxwell sector, we compute branch-resolved quasinormal-mode shifts with exterior complex scaling for the photon-sphere modes and Frobenius shooting for the near-extremal zero-damping modes. For the Schwarzschild $l=2$ fundamental mode we obtain $\beta\,\delta\omega=-0.7733-0.3946\,i$ (units $M=\alpha=1$). Off resonance, the gravitational-led shift scales as $\zeta\,l^3$ at $Q=0$ but as $\zeta\,l^2/Q$ at fixed nonzero charge and large $l$. The electromagnetic-led branch becomes exactly degenerate with the pseudoscalar mode at $Q_c=2M\sqrt{l(l+1)+1}/[l(l+1)+2]$, where the pseudoscalar potential coincides with the polar partner of the electromagnetic-led potential; there the breaking is $\mathcal{O}(\sqrt\zeta)$, $M|\Delta\omega|\simeq2.4\sqrt\zeta$ for $l=2$, and the shift reverses sign across $Q_c$. Near extremality the photon-sphere shifts remain finite, whereas the non-degenerate fundamental gravitational-led zero-damping mode is suppressed, $\delta\omega/\omega\simeq6.0\times10^{3}\,\zeta\,(1-Q^2/M^2)$ for $l=2$.
\end{abstract}

\maketitle
\thispagestyle{plain}

%======================================================================

\section{Introduction}
\label{sec:intro}

The quasinormal-mode (QNM) ringdown of a perturbed black hole encodes the
parameters of the remnant and, through the no-hair theorems, provides a
stringent test of general relativity~\cite{Kokkotas1999,Berti2009,
Konoplya2011}. With the detection of gravitational waves from binary
mergers~\cite{LIGO2016}, the ringdown phase has become an observational
probe of strong-field gravity, and the program of black-hole spectroscopy
aims to test the Kerr hypothesis by measuring multiple QNM
frequencies~\cite{Dreyer2004,Berti2006,Isi2019,Berti2018}. Any consistent modification of general
relativity generically alters the QNM spectrum, so precise predictions of
these deviations are essential to interpreting present and future ringdown
observations.

Among the best-motivated extensions is dynamical Chern--Simons (dCS)
gravity~\cite{JackiwPi2003,YunesPretorius2009,AlexanderYunes2009}, a theory with parity-violating phenomenology in which a
dynamical pseudoscalar couples to the gravitational Pontryagin density. It
arises in string theory through the Green--Schwarz anomaly-cancellation mechanism~\cite{GreenSchwarz1984,AlexanderPeskin2006}, in loop quantum gravity when the Barbero--Immirzi parameter is promoted to a field~\cite{TaverasYunes2008}, and in effective field theories of
inflation~\cite{Weinberg2008EFT}, and is the leading parity-odd curvature-squared correction to the
Einstein--Hilbert action. Solar-System tests constrain its nondynamical version~\cite{Smith2008}, and the dynamical theory is constrained by multimessenger neutron-star observations and by gravitational waves from binary mergers~\cite{Silva2021NS,Perkins2021}, and binary-black-hole coalescences have been simulated in it~\cite{Okounkova2017,Okounkova2019}. Since the Pontryagin density vanishes on any
spherically symmetric spacetime, the Schwarzschild and Reissner--Nordstr\"om
solutions persist as exact backgrounds in dCS gravity; the theory nonetheless
modifies their perturbations, and the QNM spectrum of Schwarzschild black holes
in dCS gravity has been studied in
detail~\cite{Cardoso2009,Molina2010,Kimura2018}. The rotating case has attracted recent attention: the quasinormal modes of
slowly rotating black holes in dCS gravity have been computed by direct
integration and, more recently, through modified-Teukolsky and spectral
methods~\cite{Wagle2022,Srivastava2021,Li2025,ChungLamYunes2025}.

The odd parity of the pseudoscalar has a sharp consequence for the
spectrum. In general relativity the axial (odd-parity) and polar (even-parity)
perturbations of a black hole are \emph{isospectral}: they share identical QNM
frequencies, a degeneracy guaranteed by the Chandrasekhar--Detweiler
transformation~\cite{Chandrasekhar1983}. An odd-parity field that couples to the axial sector alone lifts
this degeneracy. In dCS gravity the pseudoscalar couples only to the axial
sector, shifting the axial gravitational modes while leaving the polar sector
untouched, so that the axial and polar spectra no longer
coincide~\cite{Cardoso2009,Molina2010}. This breaking of isospectrality is a
characteristic ringdown signature of the dCS coupling, arising because the pseudoscalar perturbation has odd parity, distinct from
a mere numerical shift of individual frequencies. Isospectrality breaking has been identified as a discriminator in other
extensions of general relativity, including dilatonic, higher-curvature and $f(R)$
theories~\cite{BlazquezSalcedo2019,Bhattacharyya2017}, and is built into parametrized descriptions of the ringdown~\cite{Cardoso2019Param,McManus2019Param}; here it is driven by the odd-parity
Chern--Simons pseudoscalar and controlled by the black-hole charge.

While the Schwarzschild case is well studied, the role of black-hole
\emph{charge} has received little attention. Astrophysical black holes are expected to be nearly neutral, but charge is a useful theoretical control parameter, minicharged and dark-sector models allow sizeable charge-to-mass ratios~\cite{Cardoso2016Minicharged}, and gravitational-wave observations already constrain it~\cite{Bozzola2021,Carullo2022}. Charge also
enters the dCS problem in a nontrivial way. It couples the electromagnetic and
gravitational perturbations already in general relativity~\cite{Zerilli1974,Moncrief1974,Chandrasekhar1983,KokkotasSchutz1988,Leaver1990RN,BertiKokkotas2005}, and it deforms the very curvature invariants through which the pseudoscalar
sources the metric. Its rotating (Kerr--Newman) generalization already has a rich gravito-electromagnetic spectrum in general relativity, including eigenvalue repulsion between mode families~\cite{Pani2013KN,Dias2015KN,Dias2022KN}. The present problem should also be contrasted with our earlier study of stationary axial perturbations of the same background~\cite{RoutAltschulRN2026}, where
time-reversal symmetry forces the Chern--Simons scalar to be constant and no observable effect survives at first order in the coupling. The quasinormal modes considered here carry complex frequencies and therefore break time-reversal symmetry; the coupling is dynamical, entering at $O(\zeta)$—second order in the coupling amplitude—and is precisely the effect that the stationary analysis cannot capture.

In this work we derive the dCS couplings of the axial sector of the RN black hole and compute the resulting isospectrality breaking. Part of the calculation carries over from the Schwarzschild case: by direct reduction of the linearized field equations, the coupling of the pseudoscalar to the axial gravitational field and its reciprocal are both built from the kernel $(Mr-Q^2)\Delta/r^8$, which replaces $Mf/r^5$ and reproduces Ref.~\cite{Molina2010} at $Q=0$. We present this part compactly. The emphasis of the paper is on three features that have no Schwarzschild analog:
\begin{itemize}
\item \emph{An electromagnetic-led branch that is also shifted.} At finite charge the reduction produces a new channel: the pseudoscalar couples, in both directions, to $\Psi+\chi\Psi_e$ with $\chi=2Q/[r\sqrt{(l-1)(l+2)}]$. The electromagnetic-led quasinormal branch, absent for Schwarzschild, therefore also loses its axial--polar degeneracy. It becomes exactly degenerate with the pseudoscalar mode at $Q_c=2M\sqrt{l(l+1)+1}/[l(l+1)+2]$, where the breaking is enhanced to $\mathcal{O}(\sqrt{\zeta})$ and changes sign, an avoided crossing analogous to those of Kerr--Newman spectra~\cite{Dias2022KN}.
\item \emph{Charge controls both branches.} The Einstein--Maxwell sector decouples under a constant, $Q$-dependent rotation, so $Q$ fixes how the pseudoscalar coupling is shared between the two branches. This changes the off-resonance multipole scaling of the gravitational-led shift from $l^3$ ($Q=0$) to $l^2/Q$ (fixed $Q>0$, $Ql\gg M$), makes its charge dependence non-monotonic, and switches on the electromagnetic-led shift as $Q^2$.
\item \emph{The extremal limit.} The kernel $(Mr-Q^2)$ develops a node at the degenerate horizon. It suppresses the shift of the non-degenerate zero-damping modes~\cite{Hod2008,Yang2013,ZimmermanMark2016}, which exist only near extremality, as $1-Q^2/M^2$, while the photon-sphere modes remain unsuppressed.
\end{itemize}
We compute the shifts with an exterior-complex-scaling spectral method for the photon-sphere modes and a Frobenius--shooting method for the zero-damping family, and we state throughout which results are exact, which are numerical, and which are asymptotic estimates (Table~\ref{tab:status}).

The remainder of this paper is organized as follows. In
Sec.~\ref{sec:theory} we set up dynamical Chern--Simons gravity on the static
Reissner--Nordstr\"om background and establish the perturbation scheme, and in
Sec.~\ref{sec:parity} we show that the odd parity of the pseudoscalar confines the
Chern--Simons effect to the axial sector, breaking isospectrality. Section~\ref{sec:backbone}
records the general-relativistic axial backbone---the coupled
electromagnetic--gravitational Regge--Wheeler system of the charged black hole.
In Secs.~\ref{sec:cotton} and~\ref{sec:CdCS} we derive the axial Cotton source
and reduce it to the closed-form coupling $C_{\rm dCS}$, calibrating the
Schwarzschild limit against known results; Sec.~\ref{sec:CThetag} obtains the
reciprocal scalar-sector coupling and the electromagnetic channel, and Sec.~\ref{sec:coupled-system} assembles
the full coupled system. The isospectrality-breaking shift is analyzed in four regimes: the eikonal limit, where the branch structure is resolved (Sec.~\ref{sec:eikonal}); the low-frequency and static limit (Sec.~\ref{sec:lowfreq}); the near-extremal limit (Sec.~\ref{sec:extremal}); and weak coupling (Sec.~\ref{sec:weakcoupling}). Section~\ref{sec:numval} validates the baseline Reissner--Nordstr\"om spectrum, and Sec.~\ref{sec:wkb} presents the numerically computed shifts of the coupled system. We conclude in Sec.~\ref{sec:conclusions}. Numerical methods and convergence tests are collected in Appendix~\ref{app:numerics}.

We work in geometrized units $G=c=1$ with signature $(-,+,+,+)$. Extending the analysis to Kerr--Newman backgrounds, where rotation induces additional parity-violating effects already at first order in spin, is a natural direction for future work.

\section{Theoretical Framework}
\label{sec:theory}

\subsection{Action and Field Equations}
\label{sec:action_field_eq}
We work in dynamical Chern--Simons (dCS) gravity minimally coupled to a Maxwell
field, in the \emph{gravitational} formulation in which the pseudoscalar
$\vartheta$ couples to the gravitational Pontryagin density. The action is
\begin{equation}
\begin{aligned}
    S=\int d^4x\sqrt{-g}&\Bigg[\frac{1}{2\kappa}\,R
+\frac{\alpha}{4}\,\vartheta\,\pRR
-\frac{\beta}{2}\,g^{\mu\nu}\nabla_\mu\vartheta\,\nabla_\nu\vartheta\\
&-\frac{1}{2\kappa}\,F_{\mu\nu}F^{\mu\nu}\Bigg],
\qquad \kappa\equiv 8\pi G ,
\end{aligned}
\label{eq:action}
\end{equation}
where $F_{\mu\nu}=\partial_\mu A_\nu-\partial_\nu A_\mu$, and the Pontryagin
density is
\begin{equation}
\pRR \;=\; \dual{R}{}^{\mu}{}_{\nu}{}^{\rho\sigma}\,R^{\nu}{}_{\mu\rho\sigma},
\qquad
\dual{R}{}^{\mu}{}_{\nu}{}^{\rho\sigma}
=\tfrac{1}{2}\,\varepsilon^{\rho\sigma\tau\eta}\,R^{\mu}{}_{\nu\tau\eta},
\label{eq:pontryagin-def}
\end{equation}
with $\varepsilon^{\rho\sigma\tau\eta}$ the Levi-Civita tensor.

Varying \eqref{eq:action} with respect to the metric, the pseudoscalar, and the
gauge field gives, respectively,
\begin{align}
R_{\mu\nu}+2\kappa\alpha\,C_{\mu\nu}
&=\kappa\left(T_{\mu\nu}-\tfrac{1}{2}g_{\mu\nu}T\right),
\label{eq:einstein}\\[2pt]
\beta\,\Box\vartheta &= -\frac{\alpha}{4}\,\pRR,
\label{eq:scalar}\\[2pt]
\nabla_\mu F^{\mu\nu} &= 0,
\label{eq:maxwell}
\end{align}
where $T_{\mu\nu}=T^{\rm EM}_{\mu\nu}+T^{\vartheta}_{\mu\nu}$, and the four-dimensional generalization of the Cotton tensor (hereafter, the $C$-tensor) is
\begin{equation}
C^{\mu\nu}=(\nabla_\rho\vartheta)\,
\varepsilon^{\rho\sigma\tau(\mu}\nabla_\tau R^{\nu)}{}_{\sigma}
+(\nabla_\rho\nabla_\sigma\vartheta)\,\dual{R}{}^{\sigma(\mu\nu)\rho}.
\label{eq:ctensor}
\end{equation}
Round brackets denote symmetrization with weight $1/2$. The stress tensors are
\begin{equation}
\begin{aligned}
T^{\rm EM}_{\mu\nu}&=\frac{2}{\kappa}\Big(F_{\mu\alpha}F_{\nu}{}^{\alpha}-\tfrac14\,g_{\mu\nu}F_{\alpha\beta}F^{\alpha\beta}\Big),\\
T^{\vartheta}_{\mu\nu}&=\beta\Big(\nabla_\mu\vartheta\,\nabla_\nu\vartheta-\tfrac12\,g_{\mu\nu}\nabla_\alpha\vartheta\,\nabla^\alpha\vartheta\Big).
\end{aligned}
\label{eq:stress}
\end{equation}

\paragraph{Conventions.}
\begin{itemize}
\item \emph{Orientation.} The Levi-Civita tensor is $\varepsilon_{\mu\nu\rho\sigma}=\sqrt{-g}\,[\mu\nu\rho\sigma]$ with $[tr\theta\varphi]=+1$, so $\varepsilon^{tr\theta\varphi}=-1/\sqrt{-g}$. The dual acts on the second index pair, as in Eq.~\eqref{eq:pontryagin-def}. Reversing the orientation flips the sign of every off-diagonal Chern--Simons coupling below simultaneously ($C_{\rm dCS}$, $C_{e\Theta}$, $C_{\Theta g}$, $C_{\Theta e}$). The field redefinition $\Theta\to-\Theta$ then maps the coupled system back to itself, so the spectrum is orientation independent, including near avoided crossings.
\item \emph{Harmonics.} We work with axisymmetric modes, $Y_l\equiv Y_{l0}\propto P_l(\cos\theta)$. The background is spherically symmetric, so the radial equations do not depend on $m$ and every result holds for the whole $(2l+1)$-fold multiplet. The axial vector harmonic is $W(\theta)=\sin\theta\,\partial_\theta Y_l$ [the $S(\theta)$ of Eq.~\eqref{eq:axial-ansatz}]; it obeys $W''=\cot\theta\,W'-l(l+1)W$. Overall harmonic normalizations drop out of the master equations.
\item \emph{Shorthand.} We write $L\equiv l(l+1)$ and $\mu\equiv\sqrt{(l-1)(l+2)}$.
\end{itemize}

The Maxwell equation~\eqref{eq:maxwell} is unmodified: there is no direct $\vartheta F\tilde F$ coupling. The electromagnetic perturbation is nevertheless coupled to the pseudoscalar indirectly, through the metric perturbation and the background Coulomb field. At the level of the master equations this indirect coupling produces the electromagnetic channel derived in Sec.~\ref{sec:CThetag}.

\subsection{Background and expansion scheme}
\label{sec:background}

The static Reissner--Nordstr\"om background metric and gauge field are
\begin{equation}
ds^2=-f\,dt^2+\frac{dr^2}{f}+r^2d\Omega^2,
\qquad A_\mu dx^\mu=-\frac{Q}{r}\,dt,
\label{eq:rn-bg}
\end{equation}
with $f(r)=1-2M/r+Q^2/r^2=\Delta/r^2$ and $\Delta=(r-r_+)(r-r_-)$,
$r_\pm=M\pm\sqrt{M^2-Q^2}$. With the Maxwell normalization $-F_{\mu\nu}F^{\mu\nu}/(2\kappa)$ of Eq.~\eqref{eq:action}, the geometrized RN charge parameter $Q$ enters $f$ as $Q^2/r^2$; all occurrences of $Q$ below refer to this convention.

We expand about the RN background in two small quantities: the amplitude $\epsilon$ of the linear perturbation, and the dCS coupling. Schematically,
\begin{equation}
\begin{aligned}
g_{\mu\nu}&=g^{\rm RN}_{\mu\nu}+\epsilon\big(h^{\rm GR}_{\mu\nu}+\alpha\,h^{(1)}_{\mu\nu}+\dots\big),\\
A_\mu&=A^{\rm RN}_\mu+\epsilon\big(\delta A^{\rm GR}_\mu+\alpha\,\delta A^{(1)}_\mu+\dots\big),\\
\vartheta&=\epsilon\big(\vartheta^{(0)}+\tfrac{\alpha}{\beta}\,\vartheta^{(1)}+\dots\big).
\end{aligned}
\label{eq:expansion}
\end{equation}
Here $\vartheta^{(0)}$ is a free pseudoscalar wave on RN. It is retained because the coupled boundary-value problem contains scalar-led as well as gravitational-led modes. For the continuation of an isolated general-relativistic gravitational mode, $\vartheta^{(0)}=0$: the scalar component is then fixed by the coupled boundary conditions at $\mathcal{O}(\alpha/\beta)$ and vanishes continuously with the coupling away from a resonance. The background scalar vanishes because $\pRR|_{\rm RN}=0$. Everything below is linear in $\epsilon$. The two couplings enter as follows:
\begin{itemize}
\item the pseudoscalar drives the axial metric with a strength $\propto\alpha$, through $2\kappa\alpha C_{\mu\nu}$;
\item the metric drives the pseudoscalar with a strength $\propto\alpha/\beta$, through $-\alpha\,\pRR/4\beta$.
\end{itemize}
A shift of a gravitational quasinormal frequency requires both, so it is controlled by their product $\propto\alpha^2/\beta$, i.e.\ by the dimensionless coupling
\begin{equation}
\zeta\equiv\frac{\alpha^2}{\beta\,\kappa\,M^4}.
\label{eq:zeta}
\end{equation}
For numerical values we use units $M=\alpha=1$ (with $\kappa=8\pi$), in which $\zeta=1/(8\pi\beta)$. A shift quoted as $\beta\,\delta\omega$ therefore converts as $M\delta\omega=8\pi\,(\beta\,\delta\omega)\,\zeta$. The notations and conventions used are mentioned in Table~\ref{tab:conventions}.

\begin{table*}[t]
\caption{Notation and conventions. The last column gives $n$ in $[\text{length}]^n$ ($G=c=1$); $n=0$ is dimensionless.}
\label{tab:conventions}
\begin{tabular}{lll}
\hline\hline
Symbol & Definition & $n$\\
\hline
$M$, $Q$, $r$ & mass, charge, areal radius & $1$\\
$\omega$ & quasinormal frequency, perturbations $\propto e^{-i\omega t}$ & $-1$\\
$\kappa$ & $8\pi G$ & $0$\\
$\alpha$ & Chern--Simons coupling, $\tfrac{\alpha}{4}\vartheta\,{}^*\!RR$ in Eq.~\eqref{eq:action} & $2$\\
$\beta$ & scalar kinetic normalization & $0$\\
$\zeta$ & $\alpha^2/(\beta\kappa M^4)$ & $0$\\
$L$, $\mu$ & $l(l+1)$, $\sqrt{(l-1)(l+2)}$ & $0$\\
$\Psi_{\rm RW}$ & Regge--Wheeler function, $fh_1/r$ & $0$\\
$\Psi$ & gravitational master function, $i\Psi_{\rm RW}/\omega$ & $1$\\
$\Psi_e$ & Maxwell master function, $-2u/\mu$ with $\delta A_\varphi=u\,e^{-i\omega t}S$ & $1$\\
$\Theta$ & pseudoscalar master function, $\Theta=r\,\delta\vartheta$ per harmonic & $1$\\
$\chi$ & electromagnetic channel weight, $2Q/(r\mu)$ & $0$\\
$\beta\,\delta\omega$ & shift coefficient at $M=\alpha=1$; physical shift $M\delta\omega=8\pi\zeta\,(\beta\,\delta\omega)$ & 0\\
\hline\hline
\end{tabular}
\end{table*}

The orders that appear below are:
\begin{itemize}
\item \emph{Off resonance} (gravitational and pseudoscalar frequencies well separated): $\delta\omega=\mathcal{O}(\zeta)$.
\item \emph{Near a degeneracy} with a pseudoscalar mode: an avoided crossing gives $\delta\omega=\mathcal{O}(\sqrt{\zeta})$ (Secs.~\ref{sec:eikonal},~\ref{sec:extremal},~\ref{sec:weakcoupling}).
\item \emph{Strictly stationary perturbations}: the direct source of the Regge--Wheeler channel vanishes at $\omega=0$ (Sec.~\ref{sec:lowfreq}).
\item \emph{Pseudoscalar stress-energy}: $T^\vartheta_{\mu\nu}$ is quadratic in $\vartheta=\mathcal{O}(\epsilon)$, hence $\mathcal{O}(\epsilon^2)$. It does not enter the linear perturbation equations at any order in $\alpha$, so it can be dropped consistently even though the observable shift is itself $\mathcal{O}(\zeta)$.
\end{itemize}
The linear metric and Maxwell perturbations are therefore sourced only through the $C$-tensor~\eqref{eq:ctensor}, evaluated on the dynamical pseudoscalar and the RN background. The $\mathcal{O}(\alpha^2/\beta)$ correction $\delta V_\Theta$ to the pseudoscalar potential (Sec.~\ref{sec:coupled-system}) affects gravitational frequencies only at $\mathcal{O}(\zeta^2)$.

\section{Parity structure of the coupled perturbations}
\label{sec:parity}

The coupling $\tfrac{\alpha}{4}\vartheta\,\pRR$ is the product of two pseudoscalars and is therefore parity even; with the parity-even Einstein--Hilbert, Maxwell and scalar-kinetic terms, the action~\eqref{eq:action} is parity invariant, and the linearized equations split into multiplets of definite parity. A pseudoscalar perturbation transforms as $(-1)^{l+1}$, like the axial (odd-parity) gravitational and electromagnetic perturbations, so $\delta\vartheta$ belongs to the axial sector:
\begin{equation}
\text{polar: }\{\delta\Psi^{\rm even},\delta\phi^{\rm even}\},\qquad
\text{axial: }\{\delta\Psi^{\rm odd},\delta\phi^{\rm odd},\delta\vartheta\}.
\end{equation}
Maxwell theory is parity even, so charge adds an electromagnetic member to each sector without mixing them, and the minimal theory has no $\vartheta F\tilde F$ coupling. We have verified directly that the source of the scalar equation vanishes for a general polar (Regge--Wheeler-gauge) perturbation, $\delta(\pRR)|_{\rm polar}=0$, while it is nonzero for axial perturbations (Sec.~\ref{sec:CThetag}). The polar sector therefore coincides with that of general relativity at this order, and all Chern--Simons physics resides in the axial sector.

%======================================================================
\section{The general-relativistic axial backbone}
\label{sec:backbone}

Before introducing the Chern--Simons source, we record the general-relativistic
axial sector of the charged black hole: the coupled electromagnetic–gravitational
Regge--Wheeler system of the Reissner--Nordstr\"om (RN) spacetime. This is the
backbone onto which the dynamical Chern--Simons coupling is layered in the
following sections.

\subsection{Axial ansatz and the electromagnetic--gravitational coupling}
\label{sec:backbone-coupling}

For an odd-parity mode of harmonic index $l$ and frequency $\omega$ we work in
the Regge--Wheeler gauge, writing
\begin{equation}
\begin{aligned}
h_{t\varphi}&=h_0(r)\,e^{-i\omega t}\,S(\theta),\quad
h_{r\varphi}=h_1(r)\,e^{-i\omega t}\,S(\theta),\\
\delta A_\varphi&=u(r)\,e^{-i\omega t}\,S(\theta),\qquad
S(\theta)\equiv\sin\theta\,\partial_\theta P_l(\cos\theta),
\end{aligned}
\label{eq:axial-ansatz}
\end{equation}
with all remaining components fixed to zero by the gauge choice.

The charge couples the electromagnetic and gravitational perturbations already at
the linear level. The background Coulomb field $F_{tr}=-Q/r^2$ has its indices
raised by the axial metric perturbation, $\delta g^{\varphi t}=h_{t\varphi}/(f
r^2\sin^2\theta)$ and $\delta g^{\varphi r}=-f\,h_{r\varphi}/(r^2\sin^2\theta)$,
producing
\begin{equation}
F^{t\varphi}_{\rm coup}=-\frac{Q\,h_{r\varphi}}{r^4\sin^2\theta},\qquad
F^{r\varphi}_{\rm coup}=+\frac{Q\,h_{t\varphi}}{r^4\sin^2\theta}.
\end{equation}
Substituting into $\nabla_\mu F^{\mu\varphi}=(1/\sqrt{-g})\,\partial_\mu(\sqrt{-g}\,
F^{\mu\varphi})$ with $\sqrt{-g}=r^2\sin\theta$ yields the
electromagnetic--gravitational coupling in the axial Maxwell equation,
\begin{equation}
\big[\nabla_\mu F^{\mu\varphi}\big]_{\rm coup}
=\frac{Q}{r^{5}}\Big[r\big(h_0'+i\omega\,h_1\big)-2h_0\Big]
\,e^{-i\omega t}\,\frac{S(\theta)}{\sin^2\theta},
\label{eq:CQcomponent}
\end{equation}
which we have verified by direct computation. The common factor
$g^{\varphi\varphi}=1/(r^2\sin^2\theta)$ shared by all terms of the
$\nu=\varphi$ equation divides out, leaving a radial coupling proportional to $Q$
between the electromagnetic function $u$ and the gravitational functions
$h_0,h_1$. The reciprocal gravitational-to-electromagnetic coupling enters the
axial Einstein equation through $\delta T^{\rm EM}_{\rm axial}$, is likewise of
order $Q$, and mirrors \eqref{eq:CQcomponent} by the self-adjointness of the
linearised Einstein--Maxwell system.

\subsection{Master-variable form}
\label{sec:backbone-master}

After separation with tensor spherical harmonics and reduction to master
variables, the axial sector reduces to the coupled RN system of Chandrasekhar,
which we take in the form given by Berti and
Kokkotas~\cite{BertiKokkotas2005}. With $\Delta=r^2-2Mr+Q^2$, tortoise coordinate
$dr_*/dr=r^2/\Delta$, and master functions $H_2^{(-)}$ (gravitational) and
$H_1^{(-)}$ (electromagnetic),
\begin{align}
\left(\frac{d^2}{dr_*^2} + \omega^2\right) H_2^{(-)}
&= \frac{\Delta}{r^5}\bigg\{\Big[l(l+1)r - 6M + \frac{4Q^2}{r}\Big] H_2^{(-)}
\nonumber\\
&\quad + 2Q\sqrt{(l-1)(l+2)}\,H_1^{(-)}\bigg\},
\label{eq:H2coupled}\\[4pt]
\left(\frac{d^2}{dr_*^2} + \omega^2\right) H_1^{(-)}
&= \frac{\Delta}{r^5}\bigg\{\Big[l(l+1)r + \frac{4Q^2}{r}\Big] H_1^{(-)}
\nonumber\\
&\quad + 2Q\sqrt{(l-1)(l+2)}\,H_2^{(-)}\bigg\}.
\label{eq:H1coupled}
\end{align}
The gravitational equation carries the Regge--Wheeler term
$-6M\Delta/r^5=f(-6M/r^3)$ together with the charge term $4Q^2/r$; the
electromagnetic equation carries only the charge term. The two fields are coupled
by the off-diagonal coefficient
\begin{equation}
C_Q(r)=\frac{\Delta}{r^5}\,2Q\sqrt{(l-1)(l+2)},
\label{eq:CQ}
\end{equation}
of order $Q$, vanishing for $l=1$ (the dipole electromagnetic perturbation
being gravitationally non-radiative; the gravitational modes of interest have
$l\geq2$). The scaling $C_Q\propto Q/r^5$ of \eqref{eq:CQ} matches that of the
component-level coupling \eqref{eq:CQcomponent}, providing a consistency check
between the master-variable system and the direct calculation.\footnote{The notation $H_{2}^{(-)}$ for the gravitational wavefunction and $H_{1}^{(-)}$ for the electromagnetic  wavefunction is directly established from Berti-Kokkotas~\cite{BertiKokkotas2005}. From here on in the paper we refer to them as $H_{2}^{(-)} = \Psi$ and $H_{1}^{(-)} = \Psi_{e}$.} With this normalization $\Psi_e=-2u/\mu$, i.e.\ $\delta A_\varphi=-(\mu/2)\,\Psi_e\,e^{-i\omega t}S(\theta)$, and $\Psi_{\rm RW}=f h_1/r$.

\subsection{Chandrasekhar decoupling}
\label{sec:backbone-decouple}

A constant rotation in field space decouples
\eqref{eq:H2coupled}--\eqref{eq:H1coupled} into two Schr\"odinger-type equations
\begin{equation}
\begin{aligned}
    &\left(\frac{d^2}{dr_*^2}+\omega^2\right)Z_i^{(-)}=V_i^{(-)}Z_i^{(-)},
\\
V_i^{(-)}&=\frac{\Delta}{r^5}\Big[l(l+1)r-q_j+\frac{4Q^2}{r}\Big], \quad (i\neq j),
\end{aligned}
\label{eq:decoupled}
\end{equation}
with
\begin{equation}
q_{1,2}=3M\pm\sqrt{9M^2+4Q^2(l-1)(l+2)}.
\label{eq:qpm}
\end{equation}
The rotation is exact and independent of $r$. Write the potential matrix of Eqs.~\eqref{eq:H2coupled}--\eqref{eq:H1coupled} as $(\Delta/r^5)\big[(l(l+1)r+4Q^2/r)\,\mathbb{I}+\mathsf{A}\big]$, with $\mathbb{I}$ the $2\times2$ unit matrix. The remainder
\begin{equation}
\mathsf{A}=\begin{pmatrix}-6M & 2Q\mu\\[2pt] 2Q\mu & 0\end{pmatrix}
\label{eq:Amatrix}
\end{equation}
has constant entries, with eigenvalues $-q_{1,2}$. Define
\begin{equation}
s\equiv\sqrt{9M^2+4Q^2\mu^2},\qquad \tan\phi=\frac{s-3M}{2Q\mu}.
\end{equation}
Then
\begin{equation}
Z^{(-)}_2=\cos\phi\,\Psi-\sin\phi\,\Psi_e,\qquad
Z^{(-)}_1=\sin\phi\,\Psi+\cos\phi\,\Psi_e .
\label{eq:rotation}
\end{equation}
\begin{itemize}
\item $Z^{(-)}_2$ carries $q_1$ in Eq.~\eqref{eq:decoupled}. It is the \emph{gravitational-led} branch, continuous with the Schwarzschild Regge--Wheeler mode; we also write it $Z_g$.
\item $Z^{(-)}_1$ carries $q_2$. It is the \emph{electromagnetic-led} branch, $Z_e$.
\item The mixing angle grows from $\phi=0$ at $Q=0$ to $\phi\to\pi/4$ when $Q\mu\gg M$. At large $l$ both branches are therefore nearly equal mixtures of $\Psi$ and $\Psi_e$.
\end{itemize}
With the $i\neq j$ convention of Eq.~\eqref{eq:decoupled}, $Z_g\equiv Z^{(-)}_2$ is governed by the potential containing $q_1$ and $Z_e\equiv Z^{(-)}_1$ by the potential containing $q_2$: one checks directly that $\mathsf{A}(\cos\phi,-\sin\phi)^{T}=-q_1(\cos\phi,-\sin\phi)^{T}$ and $\mathsf{A}(\sin\phi,\cos\phi)^{T}=-q_2(\sin\phi,\cos\phi)^{T}$. As $Q\to0$, $\phi\to0$ and $q_1\to6M$, so $Z_g\to\Psi$ with the Regge--Wheeler potential, and $Z_e\to\Psi_e$. At $Q=M$, $q_1=2M(l+2)$ and $q_2=-2M(l-1)$, which give the conformal weights $h_g=l$ and $h_e=l+2$ of Sec.~\ref{sec:extremal-nh}.
The decoupled potentials~\eqref{eq:decoupled} give the pure RN spectrum. Their fundamental frequencies are the baseline against which the Chern--Simons shift is measured (Sec.~\ref{sec:numval}). The pseudoscalar couples to a fixed, $r$-dependent combination of the two branches (Sec.~\ref{sec:coupled-system}); its couplings are derived in Secs.~\ref{sec:cotton}--\ref{sec:CThetag}.

\section{The axial Chern--Simons source on static Reissner--Nordstr\"om}
\label{sec:cotton}

In dynamical Chern--Simons (dCS) gravity the pseudoscalar $\vartheta$ sources
the metric through the C-tensor (Cotton tensor). For the parity reasons
established in Sec.~\ref{sec:parity}, a dynamical pseudoscalar perturbation on a static Reissner--Nordstr\"om (RN) background produces a
nonvanishing C-tensor; this section derives the two
independent axial components that source the gravitational sector. Sections~\ref{sec:cotton}--\ref{sec:CdCS} follow the Schwarzschild calculation step by step; charge enters only through the Ricci terms of the background and the replacement $Mf/r^5\to(Mr-Q^2)\Delta/r^8$. Readers interested mainly in the results may proceed to Eq.~\eqref{eq:CdCSresult}. The genuinely new structure, the electromagnetic channel, appears in Sec.~\ref{sec:CThetag}.

\subsection{Perturbation and its covariant derivatives}
\label{sec:vderiv}

We take the dynamical scalar perturbation on the static RN background
($a=0$, no frame dragging), decomposed into scalar harmonics,
\begin{equation}
  \delta\vartheta = \sum_{l} R_l(r)\,Y_l(\theta)\,e^{-i\omega t},\qquad m=0,
  \label{eq:scalarpert}
\end{equation}
where the radial profile $R_l(r)=\Theta(r)/r$ is left unspecified here: the $C$-tensor is linear in $\delta\vartheta$, so its components can be computed for arbitrary $R_l$, and $R_l$ is fixed only later by the coupled scalar equation (Sec.~\ref{sec:CThetag}).
Writing $v_\mu \equiv \nabla_\mu\,\delta\vartheta = \partial_\mu\,\delta\vartheta$
and $v_{\mu\nu}\equiv \nabla_\mu v_\nu = \partial_\mu v_\nu
- \Gamma^\lambda{}_{\mu\nu} v_\lambda$, and stripping the common factor
$e^{-i\omega t}$ (with $\partial_t\to -i\omega$), one finds
\begin{equation}
  v_t = -i\omega R\,Y,\qquad v_r = R'\,Y,\qquad v_\theta = R\,Y',\qquad v_\phi = 0,
  \label{eq:vmu}
\end{equation}
and, using the RN Christoffel symbols,
\begin{align}
  v_{tt} &= -\omega^2 R\,Y - \tfrac{1}{2}ff'\,R'\,Y,\notag\\
  v_{tr}&=v_{rt} = -i\omega R'\,Y + \tfrac{i\omega f'}{2f}\,R\,Y, \notag\\
  v_{t\theta} &= -i\omega R\,Y', \quad v_{rr} = R''\,Y + \tfrac{f'}{2f}\,R'\,Y, \notag\\
  v_{r\theta} &= R'\,Y' - \tfrac{1}{r}R\,Y', \quad v_{\theta\theta} = R\,Y'' + r f\,R'\,Y, \label{eq:vmunu}\\
  v_{\phi\phi} &= r f \sin^2\!\theta\,R'\,Y + \sin\theta\cos\theta\,R\,Y', \notag\\
  v_{t\phi}&=v_{r\phi}=v_{\theta\phi} = 0. \notag
\end{align}
All components in \eqref{eq:vmu}--\eqref{eq:vmunu} are verified symbolically.

\subsection{C-tensor}
\label{sec:ctensordef}

We split Eq.~\eqref{eq:ctensor} as $C^{\mu\nu}=C^{\mu\nu}_{(1)}+C^{\mu\nu}_{(2)}$ (equivalent, up to normalization, to Eq.~(39) of Ref.~\cite{RoutAltschulRN2026}):
\begin{align}
  C^{\mu\nu}_{(1)} &= \tfrac12\,v_\rho\Big(
     \varepsilon^{\rho\sigma\tau\mu}\nabla_\tau R^{\nu}{}_{\sigma}
   + \varepsilon^{\rho\sigma\tau\nu}\nabla_\tau R^{\mu}{}_{\sigma}\Big),
  \label{eq:C1def}\\
  C^{\mu\nu}_{(2)} &= \tfrac12\,v_{\rho\sigma}\Big(
     {}^{*}\!R^{\sigma\mu\nu\rho} + {}^{*}\!R^{\sigma\nu\mu\rho}\Big),
  \label{eq:C2def}
\end{align}
where $C_{(1)}$ contracts the Ricci tensor and $C_{(2)}$ the (left) dual Riemann
tensor. The relevant background curvature is that of RN. The mixed Ricci tensor
is pure charge,
\begin{equation}
  R^{t}{}_{t}=R^{r}{}_{r}=-\frac{Q^2}{r^4},\qquad
  R^{\theta}{}_{\theta}=R^{\phi}{}_{\phi}=+\frac{Q^2}{r^4},
  \label{eq:RNricci}
\end{equation}
and the RN Riemann tensor is standard~\cite{Chandrasekhar1983}. Only the axial
components $C^{t\phi}$ and $C^{r\phi}$ are needed.

\subsection{Axial components}
\label{sec:axialcotton}

\paragraph{Dual-Riemann part $C_{(2)}$.}
Contracting \eqref{eq:C2def} with $v_{\sigma\tau}$, the RN Riemann structure
selects only the $(\sigma,\tau)=(r,\theta)$ and $(\theta,r)$ entries for
$C^{t\phi}_{(2)}$, and only $(t,\theta)$ and $(\theta,t)$ for $C^{r\phi}_{(2)}$
(all requisite Riemann components verified, with $\Delta$ cancelling cleanly in
the contraction):
\begin{equation}
  C^{t\phi}_{(2)} = -\frac{3(Mr-Q^2)}{r^6\sin\theta}\,v_{r\theta},
  \qquad
  C^{r\phi}_{(2)} = \frac{3(Mr-Q^2)}{r^6\sin\theta}\,v_{t\theta}.
  \label{eq:C2results}
\end{equation}

\paragraph{Ricci part $C_{(1)}$.}
Because the mixed Ricci difference $R^{t}{}_{t}-R^{\phi}{}_{\phi}$ in
\eqref{eq:RNricci} is a function of $r$ only, its angular ($\nabla_\theta$) and
temporal ($\nabla_t$) gradients vanish; only the radial gradient survives in
$C^{t\phi}_{(1)}$, and $C^{r\phi}_{(1)}$ vanishes identically:
\begin{equation}
  C^{t\phi}_{(1)} = -\frac{3Q^2}{r^7}\,R_l\,S^\phi,
  \qquad
  C^{r\phi}_{(1)} = 0,
  \qquad S^\phi\equiv\frac{\partial_\theta Y_l}{\sin\theta}.
  \label{eq:C1results}
\end{equation}
Consistently, $C^{t\phi}_{(1)}\to 0$ as $Q\to 0$, since Schwarzschild is
Ricci-flat.

\paragraph{Combined source.}
Inserting $v_{r\theta}=(R_l'-R_l/r)\,\partial_\theta Y_l$ and $v_{t\theta}=-i\omega R_l\,\partial_\theta Y_l$ from Eq.~\eqref{eq:vmunu} into Eqs.~\eqref{eq:C2results}--\eqref{eq:C1results} and adding, both axial components carry the common angular factor $S^\phi$:
\begin{align}
C^{t\phi} &= \left[ -\frac{3(Mr-Q^2)}{r^6}\, R_l' + \frac{3(Mr-2Q^2)}{r^7}\, R_l \right] S^\phi, \label{eq:Ctphi_final} \\[6pt]
C^{r\phi} &= -\frac{3i\omega(Mr-Q^2)}{r^6}\, R_l\, S^\phi. \label{eq:Crphi_final}
\end{align}
  
Thus $C^{t\phi}$ has real radial coefficient functions with two radial channels $\{R_l',R_l\}$, while $C^{r\phi}$ carries an explicit factor $i\omega$ --- the hallmark of the dynamical ($e^{-i\omega t}$, complex $\omega$) coupling that is absent in the stationary, time-reversal-symmetric analysis. In the neutral limit both reduce to their pure-mass, Schwarzschild--dCS forms, sourced entirely by $C_{(2)}$.
Equations \eqref{eq:Ctphi_final}--\eqref{eq:Crphi_final} are the input to the master-variable reduction of Sec.~\ref{sec:CdCS}.

The contravariant components are proportional to $S^\phi$, so the covariant components $C_{\mu\phi}=g_{\mu\mu}g_{\phi\phi}C^{\mu\phi}$ are proportional to $W$. This is what makes the projection onto the gravitational sector diagonal in $l$ (Sec.~\ref{sec:CdCS}).

\section{The axial master equation and the dynamical Chern--Simons coupling}
\label{sec:CdCS}

We now reduce the axial sector to a single Schr\"odinger-type master equation
and extract the coupling coefficient $C_{\rm dCS}$ through which the dynamical
scalar drives the gravitational field. The strategy is: (i) obtain the two
first-order axial field equations, (ii) project the Cotton source of
Sec.~\ref{sec:cotton} onto the axial harmonic, (iii) eliminate the
non-dynamical metric function and pass to the Regge--Wheeler master variable,
and (iv) fix the normalisation and validate the whole construction against the
known Schwarzschild limit before reading off the Reissner--Nordstr\"om (RN)
result.

\subsection{Axial field equations}
\label{sec:CdCS-fieldeqs}

In Regge--Wheeler gauge the odd-parity metric perturbation is carried by two
radial functions $h_0(r)=h_{t\phi}$, $h_1(r)=h_{r\phi}$, both multiplying the
axial vector harmonic $W(\theta)\equiv\sin\theta\,\partial_\theta Y_l$ with time
dependence $e^{-i\omega t}$. The linearised Einstein tensor on the static RN
background gives, after using the harmonic identity
$W''=\cot\theta\,W'-L\,W$ (which separates both components cleanly, the $W'$
remainders cancelling identically),
\begin{align}
  \delta G_{t\phi}
  &= W \Bigg[ -\frac{\Delta}{2r^2} h_0'' - \frac{i\omega\Delta}{2r^2} h_1' \nonumber \\
  &\quad + \frac{Lr^2 - 4Mr + 4Q^2}{2r^4} h_0 - \frac{i\omega\Delta}{r^3} h_1 \Bigg], \label{eq:dGtphi} \\[6pt]
  \delta G_{r\phi}
  &= W \left[ \frac{i\omega r^2}{2\Delta} h_0' - \frac{i\omega r}{\Delta} h_0 + \frac{\mathcal{N}(r)}{2r^4\Delta} h_1 \right]. \label{eq:dGrphi}
\end{align}
with
\begin{equation}
  \mathcal{N}(r)=L\,r^2\Delta+4Mr(r^2-Q^2)-2(r^2-Q^2)(r^2+Q^2)-\omega^2 r^6 .
  \label{eq:Ndef2}
\end{equation}
The field equation~\eqref{eq:einstein} turns these into sourced equations,
$\delta G_{\mu\phi}=-2\kappa\alpha\,C_{\mu\phi}+\kappa\,\delta T^{\rm EM}_{\mu\phi}$ (with $2\kappa\alpha=16\pi$ at $\alpha=1$). This uses that the background has $R=0$ and $\delta R=0$ for axial perturbations, and that $T^\vartheta$ is $\mathcal{O}(\epsilon^2)$. The electromagnetic term produces the coupling $C_Q$ of Sec.~\ref{sec:backbone}.

\subsection{Angular projection: diagonal coupling}
\label{sec:CdCS-angular}

A key point fixes the whole calculation. The contravariant Cotton components of Sec.~\ref{sec:cotton} are proportional to $S^\phi=\partial_\theta Y_l/\sin\theta$, so that on lowering the index
$C_{\mu\phi}=g_{\mu\mu}g_{\phi\phi}C^{\mu\phi}\propto W$, exactly matching the
gravitational operators \eqref{eq:dGtphi}--\eqref{eq:dGrphi}. Because the
background is spherically symmetric, the map
$\delta\vartheta\!\mapsto\!C_{\mu\nu}[\delta\vartheta]$ is $SO(3)$-equivariant
and cannot mix multipoles; projecting onto $W$ with the Thorne normalisation
$\int \bar S^A S_A\,d\Omega=l(l+1)$ is therefore diagonal in $l$, the weight
$l(l+1)$ cancelling against that of the gravitational operator. The coupling is
thus a single radial function per multipole, with no $l\to l\pm2$ banding.
Writing the source coefficients (of $W$) from the verified Cotton components,
\begin{align}
  -16\pi\,C_{t\phi}\big|_W
  &= -\frac{48\pi f}{r^5}\Big[r(Mr-Q^2)R_l' - (Mr-2Q^2)R_l\Big],
  \label{eq:StRaw}\\
  -16\pi\,C_{r\phi}\big|_W
  &= \frac{48\pi\,i\omega\,(Mr-Q^2)}{f\,r^4}\,R_l .
  \label{eq:SrRaw}
\end{align}

\subsection{Elimination, master variable, and calibration}
\label{sec:CdCS-reduce}

Equation \eqref{eq:dGrphi} is first order in $h_0$; differentiating it and
inserting the result into \eqref{eq:dGtphi} renders $h_0$ algebraic, and
back-substitution yields a single second-order equation for
$h_1$. In terms of the Regge--Wheeler master variable
\begin{equation}
  \Psi_{\mathrm{RW}} \equiv \frac{\Delta}{r^3}\,h_1 ,
\end{equation}
this equation takes the form of a Schr\"odinger type master equation. In the neutral limit the reduction was
verified to reproduce the Schwarzschild Regge--Wheeler potential
$f\big[l(l+1)/r^2-6M/r^3\big]$ to all orders, confirming the elimination
combination independently of any assumed coefficients. For the coupled RN
system the gravitational potential is
$V_{\rm RW}=(\Delta/r^5)[\,l(l+1)r-6M+4Q^2/r\,]$, the electromagnetic
back-reaction residing in the separate coupling $C_Q$ to the Maxwell master
function $\Psi_{e}$.

Following the treatment of the Schwarzschild case, we adopt the axial master
function $\Psi\equiv i\Psi_{\mathrm{RW}}/\omega$. The factor $i/\omega$ is not cosmetic: it
absorbs the $\omega$-dependence of the source \eqref{eq:SrRaw}, converting what
would otherwise be a first-derivative coupling into a purely algebraic one.
The master equation becomes
\begin{equation}
  \left[\frac{d^2}{dr_*^2}+\omega^2 - V_{\rm RW}\right]\Psi
  = C_Q\,\Psi_{e} + C_{\rm dCS}\,\Theta ,
  \label{eq:mastereqfinal}
\end{equation}
where the scalar radial function $R_l$ of \eqref{eq:scalarpert} enters through $\Theta = rR_l$ and ties them. In assembling the
coefficient of $\Theta$ one uses the scalar master equation
$\big[d^2/dr_*^2+\omega^2-V_\Theta\big]\Theta=0$, with
$V_\Theta=(\Delta/r^5)[\,l(l+1)r+2M-2Q^2/r\,]$, to eliminate $\Theta''$
on shell.

\subsection{The coupling coefficient}
\label{sec:CdCS-result}

Carrying out the reduction, the first-derivative ($\Theta'$) contribution
cancels identically for all $Q$, and the coupling is algebraic:
\begin{equation}
  C_{\rm dCS}(r) = \frac{96\pi\,\alpha\,(Mr-Q^2)\,\Delta}{r^8} = \frac{96\pi\,\alpha\,(Mr-Q^2)\,f}{r^6}.
  \label{eq:CdCSresult}
\end{equation}
The coupling is built from the same combination $(Mr-Q^2)$ that appears in every
axial Cotton component, multiplied by the horizon factor $\Delta$; it carries no
explicit $l$-dependence. Several checks fix confidence in
\eqref{eq:CdCSresult}:

\begin{itemize}

\item \emph{Schwarzschild limit.} As $Q\to0$, $C_{\rm dCS}\to 96\pi\alpha M f/r^5$,
  the coupling of Ref.~\cite{Molina2010}.
\item \emph{Algebraic character.} The coupling multiplies $\Theta$ with no
  derivative term, as in the Schwarzschild case; this holds only in the
  $\Psi=i\Psi_{\rm RW}/\omega$ normalisation.
\item \emph{Horizon regularity.} The factor $\Delta$ guarantees the source
  vanishes at the outer horizon $r_+$.
\item \emph{Multipole structure.} As in the neutral case, the angular momentum
  factors $\sqrt{(l-1)l(l+1)(l+2)}$ do not appear in this gravitational-sector
  coupling; they enter the reciprocal (scalar-sector) coupling, where $l=1$ is
  non-radiative and drops out.
\end{itemize}

The Schwarzschild comparison can be made complete. Setting $Q\to0$, our coupled
system \eqref{eq:coupled}--\eqref{eq:allcouplings} reduces to
\begin{align}
\left[\frac{d^2}{dr_*^2}+\omega^2-f\Big(\tfrac{l(l+1)}{r^2}-\tfrac{6M}{r^3}\Big)\right]\Psi
&=\frac{96\pi M f}{r^5}\,\alpha\,\Theta,
\label{eq:molina-grav}\\
\left[\frac{d^2}{dr_*^2}+\omega^2-f\Big(\tfrac{l(l+1)}{r^2}+\tfrac{2M}{r^3}\Big)\right]\Theta
&=\frac{6M\alpha}{\beta}\frac{(l+2)!}{(l-2)!}\frac{f}{r^5}\,\Psi,
\label{eq:molina-scal}
\end{align}
which coincides term by term with Eqs.~(2.8)--(2.9) of Ref.~\cite{Molina2010} once the $\mathcal{O}(\alpha^2/\beta)$ back-reaction term $\delta V_\Theta$ (Sec.~\ref{sec:coupled-system}) is restored in the scalar potential. They have the same gravitational coupling $96\pi\alpha Mf/r^5$, the reciprocal coupling with
its full multipole weight, and the master-variable normalisation
$\Psi=i\Psi_{\rm RW}/\omega$. This is an independent, closed-form check of our
coupling derivation in the neutral limit; the charge dependence encoded in the
common kernel $(Mr-Q^2)\Delta/r^8$ is the new result of the present work.

The same elimination also feeds the pseudoscalar into the Maxwell equation, because the master equation for $\Psi_e$ is assembled with $h_0$ eliminated through the sourced Einstein equations. This channel and the reciprocal (scalar-sector) couplings are derived in Sec.~\ref{sec:CThetag}. Together with $C_Q$ they close the three-field axial system $\{\Psi,\Psi_e,\Theta\}$ of Sec.~\ref{sec:coupled-system}.

\section{The scalar-sector source, the reciprocal coupling, and the electromagnetic channel}
\label{sec:CThetag}

Consistency of the coupled system requires the reciprocal channel, in which the axial perturbation sources the pseudoscalar through
\begin{equation}
  \Box\,\vartheta = -\frac{\alpha}{4\beta}\,\pRR .
  \label{eq:scalarEOM}
\end{equation}
On the static RN background $\pRR=0$, so the source is the linear variation $\delta(\pRR)$ evaluated on the axial perturbation. We derive it directly; details are in Appendix~\ref{app:reciprocal}. Unlike the gravitational coupling of Sec.~\ref{sec:CdCS}, the result contains a channel with no Schwarzschild analog: at finite charge the axial Einstein equation mixes $h_0,h_1$ with the Maxwell perturbation, so the pseudoscalar is sourced by $\Psi_e$ as well as by $\Psi$.

\subsection{The linearized Pontryagin density}
\label{sec:CThetag-pont}

For the Regge--Wheeler-gauge perturbation~\eqref{eq:axial-ansatz}, a direct computation gives
\begin{equation}
  \delta(\pRR)=-\frac{24L\,(Mr-Q^2)}{r^7}\Big[r\,h_0'-2h_0+i\omega r\,h_1\Big]\,Y_l ,
  \label{eq:dstarRR}
\end{equation}
exactly, at all $Q$; all $\partial_\theta Y_l$ terms cancel. The kernel $(Mr-Q^2)$ is no accident. The Pontryagin density depends on the curvature only through the Weyl tensor,
\begin{equation}
{}^{*}\!R^{\mu}{}_{\nu}{}^{\rho\sigma}R^{\nu}{}_{\mu\rho\sigma}={}^{*}C^{\mu}{}_{\nu}{}^{\rho\sigma}C^{\nu}{}_{\mu\rho\sigma},
\label{eq:RR-CC}
\end{equation}
because in four dimensions every term containing the Ricci tensor or scalar vanishes in this parity-odd contraction by the symmetries of $\varepsilon$~\cite{AlexanderYunes2009}. The RN background is of Petrov type D, with $\Psi_2=-(Mr-Q^2)/r^4$ its only nonvanishing Weyl scalar. Hence at linear order $\delta(\pRR)\propto\Psi_2\,{\rm Im}\,\delta\Psi_2$, where ${\rm Im}\,\delta\Psi_2$ is gauge invariant at first order.

\subsection{Reduction to master variables}
\label{sec:CThetag-reduce}

At the order required, $h_0$ and $h_1$ obey the general-relativistic Einstein--Maxwell equations. The $r\phi$ component gives
\begin{equation}
  h_0'-\frac{2h_0}{r}+i\omega h_1=\frac{(L-2)}{r}\big[\Psi+\chi\,\Psi_e\big],
  \qquad
  \chi(r)\equiv\frac{2Q}{r\mu},
  \label{eq:rphi-reduced}
\end{equation}
with $\Psi=i\Psi_{\rm RW}/\omega$ and $\Psi_e$ normalized as in Eqs.~\eqref{eq:H2coupled}--\eqref{eq:H1coupled}. The $\Psi_e$ term comes from the electromagnetic stress $\delta T^{\rm EM}_{r\phi}\propto Q\,\omega\,\delta A_\phi$. Substituting into Eq.~\eqref{eq:dstarRR},
\begin{equation}
\begin{aligned}
    \delta(\pRR)&=-\frac{24\,(l-1)l(l+1)(l+2)\,(Mr-Q^2)}{r^7}\\
    &\times\big[\Psi+\chi\Psi_e\big]\,Y_l .
\end{aligned}
\label{eq:dstarRR-master}
\end{equation}
The tensor weight $(l-1)l(l+1)(l+2)=L(L-2)$ arises directly: one factor of $L$ comes from the harmonic algebra, the factor $L-2=\mu^2$ from the Einstein equation. With $\delta\vartheta=(\Theta/r)Y_l\,e^{-i\omega t}$, Eq.~\eqref{eq:scalarEOM} becomes
\begin{equation}
  \Big[\frac{d^2}{dr_*^2}+\omega^2-V_\Theta\Big]\Theta
  =C_{\Theta g}\,\Psi+C_{\Theta e}\,\Psi_e ,
\end{equation}
with
\begin{equation}
\begin{aligned}
    C_{\Theta g}&=\frac{6\alpha(Mr-Q^2)(l-1)l(l+1)(l+2)\,\Delta}{\beta\,r^8},
  \\
  C_{\Theta e}&=\chi\,C_{\Theta g}.
\end{aligned}
\label{eq:CThetag-result}
\end{equation}

\subsection{The electromagnetic channel on the gravity side}
\label{sec:CThetag-em}

The Maxwell equation is unmodified, but it involves $h_0$. Eliminating $h_0$ through the \emph{sourced} Einstein equations carries the $C$-tensor source into the Maxwell master equation:
\begin{equation}
  \Big[\frac{d^2}{dr_*^2}+\omega^2-V_{\rm EM}\Big]\Psi_e=C_Q\Psi+C_{e\Theta}\,\Theta,
  \quad
  C_{e\Theta}=\chi\,C_{\rm dCS}.
  \label{eq:CeTheta}
\end{equation}
Thus the pseudoscalar couples to the single combination $\Psi+\chi\Psi_e$ in \emph{both} directions, and
\begin{equation}
  \frac{C_{\Theta g}}{C_{\rm dCS}}=\frac{C_{\Theta e}}{C_{e\Theta}}=\frac{(l-1)l(l+1)(l+2)}{16\pi\beta}.
  \label{eq:CThetag-rel}
\end{equation}
This is the reciprocity required by the single bilinear interaction $\tfrac{\alpha}{4}\int\sqrt{-g}\,\delta\vartheta\,\delta(\pRR)$; here it is derived rather than assumed.

Three checks support these results. First, the unused $t\phi$ Einstein equation, including its $C$-tensor source, is satisfied identically on shell. Second, the $Q\to0$ limit reproduces Ref.~\cite{Molina2010}: $\chi=0$, $C_{\Theta g}\to(6M\alpha/\beta)\,[(l+2)!/(l-2)!]\,f/r^5$ [their Eq.~(2.9)]. Third, $C_{\Theta g}$ vanishes at $l=1$, as it must for the non-radiative dipole. The electromagnetic channel $\chi\propto Q/(r\mu)$ is absent in the neutral limit and suppressed as $1/l$ at large multipole.

\section{The coupled axial system}
\label{sec:coupled-system}

Collecting the gravitational (Sec.~\ref{sec:CdCS}), electromagnetic
(Sec.~\ref{sec:backbone}) and scalar (Sec.~\ref{sec:CThetag}) sectors, the
axial perturbations of the charged black hole in dCS gravity obey the
three coupled Schr\"odinger-type equations
\begin{subequations}
\label{eq:coupled}
\begin{align}
  \left[\frac{d^2}{dr_*^2}+\omega^2 - V_{\rm RW}\right]\Psi
    &= C_Q\,\Psi_e + C_{\rm dCS}\,\Theta ,
  \label{eq:coupled-g}\\[3pt]
  \left[\frac{d^2}{dr_*^2}+\omega^2 - V_{\rm EM}\right]\Psi_e
    &= C_Q\,\Psi + \chi\,C_{\rm dCS}\,\Theta ,
  \label{eq:coupled-em}\\[3pt]
  \left[\frac{d^2}{dr_*^2}+\omega^2 - V_{\Theta}\right]\Theta
    &= C_{\Theta g}\,\big(\Psi+\chi\,\Psi_e\big) ,
  \label{eq:coupled-s}
\end{align}
\end{subequations}
where $\Psi = i\Psi_{\rm RW}/\omega$ is the axial gravitational master function (Sec.~\ref{sec:CdCS}), $\Psi_e$ the Maxwell master function, $\Theta$ the pseudoscalar, and $\chi=2Q/(r\mu)$. The potentials are
\begin{align}
  V_{\rm RW} &= \frac{\Delta}{r^5}\Big[l(l+1)r-6M+\frac{4Q^2}{r}\Big],\\
  V_{\rm EM} &= \frac{\Delta}{r^5}\Big[l(l+1)r+\frac{4Q^2}{r}\Big],
  \label{eq:pots1}\\
  V_{\Theta} &= \frac{\Delta}{r^5}\Big[l(l+1)r+2M-\frac{2Q^2}{r}\Big]
    + \delta V_\Theta ,
  \label{eq:pots2}
\end{align}
and the couplings, in units $\alpha=1$ (as in Ref.~\cite{Molina2010}, $\alpha$ can be absorbed by $\beta\to\alpha^2\beta$, $\Theta\to\Theta/\alpha$, leaving only $\zeta$; restoring it, $C_{\rm dCS}\propto\alpha$ and $C_{\Theta g}\propto\alpha/\beta$), are
\begin{equation}
\begin{aligned}
C_Q &= \frac{2Q\mu\,\Delta}{r^5},\\
C_{\rm dCS} &= \frac{96\pi(Mr-Q^2)\Delta}{r^8},\\
C_{\Theta g} &= \frac{6(Mr-Q^2)(l-1)l(l+1)(l+2)\Delta}{\beta\,r^8}.
\end{aligned}
\label{eq:allcouplings}
\end{equation}
The term $\delta V_\Theta=\mathcal{O}(\alpha^2/\beta)$ is the dCS back-reaction on the pseudoscalar potential, $\delta V_\Theta=576\pi\,l(l+1)\,\alpha^2(Mr-Q^2)^2\Delta/(\beta r^{12})$, which reduces to the result of Ref.~\cite{Molina2010} at $Q=0$. It enters the gravitational shift only at $\mathcal{O}(\zeta^2)$ and is dropped.

\subsection{Branch structure}
\label{sec:branches}

In the branch basis~\eqref{eq:rotation} the general-relativistic block is diagonal. The pseudoscalar couples to $Z_g\equiv Z^{(-)}_2$ and $Z_e\equiv Z^{(-)}_1$ through the combination
\begin{equation}
\begin{aligned}
\Psi+\chi\Psi_e&=w_g(r)\,Z_g+w_e(r)\,Z_e,\\
w_g&=\cos\phi-\chi\sin\phi,\qquad w_e=\sin\phi+\chi\cos\phi .
\end{aligned}
\label{eq:weights-branch}
\end{equation}
Away from degeneracies, and after this branch projection, the two branches communicate only through $\Theta$, and only at fourth order in the couplings. The leading off-resonant shift of an isolated branch $a\in\{g,e\}$ is therefore given exactly, through $\mathcal{O}(\zeta)$, by the two-field system
\begin{equation}
\begin{aligned}
\Big[\frac{d^2}{dr_*^2}+\omega^2-V_a\Big]Z_a&=w_a\,C_{\rm dCS}\,\Theta,\\
\Big[\frac{d^2}{dr_*^2}+\omega^2-V_\Theta\Big]\Theta&=w_a\,C_{\Theta g}\,Z_a,
\end{aligned}
\label{eq:reduced}
\end{equation}
with $V_g$ and $V_e$ the potentials of Eq.~\eqref{eq:decoupled} carrying $q_1$ and $q_2$ respectively. We have checked numerically that Eq.~\eqref{eq:reduced} reproduces the $\mathcal{O}(\zeta)$ shift of the full system~\eqref{eq:coupled} to $10^{-7}$ (Appendix~\ref{app:numerics}). Near a branch--scalar degeneracy (Secs.~\ref{sec:eikonal-em} and~\ref{sec:extremal-shift}) the resonant coupled block must be retained instead. In the spectral and shooting calculations the full radial dependence of $w_a(r)$ is kept; only the local analysis of Sec.~\ref{sec:eikonal} evaluates it near the potential peak. The polar sector is unaffected, so we define the isospectrality-breaking shift branch by branch,
\begin{equation}
\Delta\omega^{(a)}_{\rm iso}\equiv\omega^{\rm axial}_{a}(\alpha,Q)-\omega^{\rm polar}_{a}(0,Q)=\omega_a(\alpha,Q)-\omega_a(0,Q),
\label{eq:iso-def}
\end{equation}
where each branch is continued from its general-relativistic ($\alpha\to0$) limit. Both the gravitational-led \emph{and} the electromagnetic-led branches are shifted.

\section{Eikonal limit: the isospectrality-breaking shift}
\label{sec:eikonal}

Integrating out $\Theta$ in Eq.~\eqref{eq:reduced} in the local approximation, appropriate at large $l$ where the modes localize at the potential peak, gives effective eigen-potentials for each branch:
\begin{equation}
\begin{aligned}
&V^{(a)}_\pm=\frac{V_a+V_\Theta}{2}\pm\sqrt{D_a^2+w_a^2P},\\
&D_a\equiv\frac{V_a-V_\Theta}{2}=-\frac{\Delta}{2r^5}\Big[2M-\frac{6Q^2}{r}+q_{(a)}\Big],\\
&P\equiv C_{\rm dCS}C_{\Theta g}=\frac{576\pi\,(l-1)l(l+1)(l+2)(Mr-Q^2)^2\Delta^2}{\beta\,r^{16}},
\end{aligned}
\label{eq:eigenpot}
\end{equation}
with $q_{(g)}=q_1$ and $q_{(e)}=q_2$. At $Q=0$, $D_g=-4M\Delta/r^5$. Equations~\eqref{eq:eigenpot}--\eqref{eq:Qc} provide a local, eikonal interpretation of the branch shifts. The finite-$l$, finite-charge coefficients of Sec.~\ref{sec:wkb} are obtained instead from the nonlocal coupled spectral eigenproblem.

\subsection{Off-resonance regime}
\label{sec:eikonal-pert}

For $w_a^2P\ll D_a^2$,
\begin{equation}
\delta(\omega_a^2)\simeq\frac{w_a^2\,P}{2D_a}\bigg|_{r_{\rm peak}},
\quad
\delta\omega_a=\frac{\delta(\omega_a^2)}{2\omega_a},
\label{eq:shift-pert}
\end{equation}
which is $\mathcal{O}(\zeta)$. Against the exact numerical coefficients of Sec.~\ref{sec:wkb} it is accurate to $\simeq11\%$ at $l=2$, $\le2\%$ at $l=4$, and $\le1\%$ at $l=12$, for $Q/M=0$, $0.3$ and $0.6$. Two scalings follow.

\paragraph{Neutral hole.} At $Q=0$ the spacing $D_g$ is independent of $l$ while $P\propto l^4$. With $r_{\rm ph}=3M$ and $\omega_g\simeq l/(3\sqrt3 M)$,
\begin{equation}
M\,\delta\omega_g\big|_{Q=0}\simeq-\frac{32\sqrt3\,\pi^2}{243}\,\zeta\,l^3\approx-2.25\,\zeta\,l^3 .
\label{eq:domega-schw}
\end{equation}

\paragraph{Charged hole.} For $Q>0$ the Einstein--Maxwell splitting grows with $l$: once $Q\mu\gtrsim\tfrac32M$, $V_\Theta-V_g\simeq2Q\mu\,\Delta/r^5$ and $w_g^2\to\tfrac12$. Then
\begin{equation}
\delta\omega_g\simeq-\frac{72\pi\,l^2}{\beta\,Q}\,\frac{K^2r^6}{\sqrt f\,\Delta}\bigg|_{r_{\rm ph}(Q)}\propto\frac{l^2}{Q},
\quad K\equiv\frac{(Mr-Q^2)\Delta}{r^8},
\label{eq:domega-Q}
\end{equation}
in units $\alpha=1$, with $r_{\rm ph}$ given by
\begin{equation}
  r_{\rm ph}=\frac{3M+\sqrt{9M^2-8Q^2}}{2}.
  \label{eq:rph}
\end{equation}

\begin{figure*}[htbp]
\centering
\includegraphics[width=\textwidth]{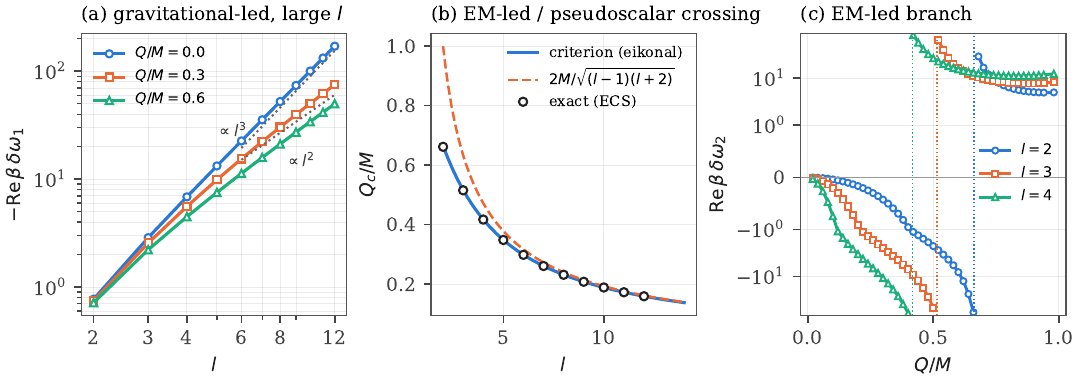}
\caption{Branch-resolved isospectrality breaking (units $M=\alpha=1$; physical shift $=$ plotted value$/\beta$).
(a)~Gravitational-led shift versus multipole: $\propto l^3$ at $Q=0$ and $\propto l^2$ at fixed $Q>0$ with $Ql\gg M$ [Eqs.~\eqref{eq:domega-schw},~\eqref{eq:domega-Q}].
(b)~Charge $Q_c$ at which the electromagnetic-led branch crosses the pseudoscalar mode: numerical crossing (circles), which coincides with the exact result~\eqref{eq:Qc}; the local criterion~\eqref{eq:Qc-local} (solid); and the large-$l$ form $2M/\sqrt{(l-1)(l+2)}$ (dashed).
(c)~Shift of the electromagnetic-led branch, which diverges at first order and reverses sign at $Q_c$ (dotted). The divergent curves are the off-resonant first-order approximation and are not physical within the resonance windows; there the finite $\mathcal{O}(\sqrt\zeta)$ avoided crossing of the coupled system applies (Sec.~\ref{sec:eikonal-em}).}
\label{fig:branches}
\end{figure*}

The limits $Q\to0$ and $l\to\infty$ do not commute. The numerical shifts confirm the change of scaling: between $l=11$ and $12$ the local exponent $d\ln|\mathrm{Re}\,\delta\omega_g|/d\ln l$ is $2.91$ at $Q=0$, but $2.20$ at $Q/M=0.3$ and $2.10$ at $Q/M=0.6$ [Fig.~\ref{fig:branches}(a)]. This change of scaling is also the origin of the strong reduction of the shift at intermediate charge for $l\geq3$ (Sec.~\ref{sec:wkb}).

\begin{figure}[htbp]
\centering
\includegraphics[width=\columnwidth]{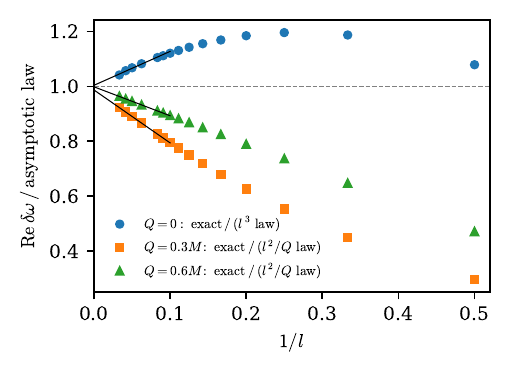}
\caption{Ratio of the spectral gravitational-led coefficient $\mathrm{Re}\,\delta\omega_g$ to the asymptotic laws: $M\delta\omega\simeq-2.25\,\zeta\,l^3$ at $Q=0$ [Eq.~\eqref{eq:domega-schw}] and the $l^2/Q$ law of Eq.~\eqref{eq:domega-Q} at $Q=0.3M,\,0.6M$, for $l=2$--$30$. Lines: linear fits in $1/l$ over $l\ge12$, extrapolating to $1.004$, $0.987$ and $0.999$.}
\label{fig:eikratio}
\end{figure}

Extending the reduced spectral calculation to $l=30$ confirms both laws quantitatively (Fig.~\ref{fig:eikratio}). The ratio of $\mathrm{Re}\,\delta\omega_g$ to Eq.~\eqref{eq:domega-schw} behaves as $1+1.24/l$ at $Q=0$, while the ratio to Eq.~\eqref{eq:domega-Q} behaves as $1-0.6M/(Ql)$ at $Q=0.3M$ and $0.6M$; linear extrapolation in $1/l$ from $l=12$--$30$ gives $1.004$, $0.987$ and $0.999$, respectively. The $l^2/Q$ law therefore requires $Ql\gg M$: taking $l\to\infty$ at fixed $Q>0$ and $Q\to0$ at fixed $l$ do not commute, and Eq.~\eqref{eq:domega-schw} is recovered only for $Ql\ll M$. Both expressions are derived asymptotic results; the quantitative coefficients are those of the full spectral calculation.

\begin{figure*}[htbp]
\centering
\includegraphics[width=\textwidth]{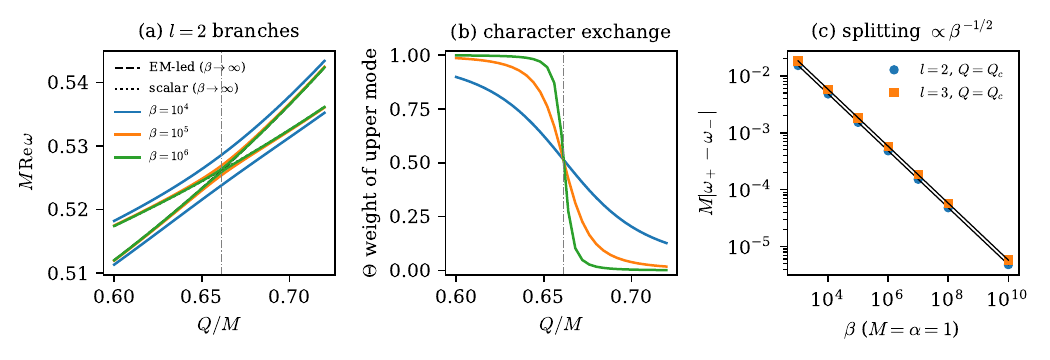}
\caption{Avoided crossing of the electromagnetic-led and pseudoscalar modes ($l=2$; units $M=\alpha=1$).
(a)~$\mathrm{Re}\,\omega$ of the two coupled modes through $Q_c$ [Eq.~\eqref{eq:Qc}, vertical line] for $\beta=10^4,10^5,10^6$; dashed and dotted curves are the decoupled ($\beta\to\infty$) modes, which cross exactly at $Q_c$.
(b)~Pseudoscalar weight of the upper mode (eigenvector norm in the $\Theta$ block, rescaled by $\sqrt{16\pi\beta/[L(L-2)]}$ to symmetrize the couplings); the two modes exchange character across a window of width $\propto\beta^{-1/2}$.
(c)~Splitting at $Q=Q_c$ for $l=2,3$: $M|\omega_+-\omega_-|\propto\beta^{-1/2}$ over seven decades (lines), i.e.\ $\mathcal{O}(\sqrt\zeta)$.}
\label{fig:avoided}
\end{figure*}

\subsection{Strong-mixing regime and the validity of the effective theory}
\label{sec:eikonal-strong}

The off-resonance regime requires $w_a^2P\ll D_a^2$. This means $\zeta l^4\ll1$ at $Q=0$, and $\zeta\,l^2\ll(Q/M)^2$ at fixed $Q>0$ with $Ql\gg M$. Beyond it, $\delta(\omega_a^2)\simeq-|w_a|\sqrt{P}$, so $\delta\omega_a\propto l\sqrt{\zeta}$: the frequency shift is $\mathcal{O}(\sqrt\zeta)$ and grows linearly with $l$. At $Q=0$ (units $\alpha = 1$) this gives
\begin{equation}
\delta\omega_g\big|_{Q=0}\simeq-\frac{4\sqrt{3\pi}}{81}\,\frac{l}{M^3\sqrt{\beta}}.
\label{eq:shift-eikonal}
\end{equation}
At fixed small $\zeta$ the perturbative result becomes increasingly accurate as $\zeta\to0$. For any fixed nonzero coupling, however, sufficiently large $l$ leaves the perturbative mixing regime. Very large $l$ also probes short scales, where dCS gravity, as an effective theory, ceases to be reliable.

\subsection{The electromagnetic-led branch: resonance and sign reversal}
\label{sec:eikonal-em}

For the electromagnetic-led branch, $q_2=3M-s$ decreases with $Q$, so $D_e$ changes sign, and the branch becomes degenerate with the pseudoscalar mode. The local criterion $D_e(r_{\rm ph})=0$,
\begin{equation}
2M-\frac{6Q^2}{r_{\rm ph}}+3M-\sqrt{9M^2+4Q^2\mu^2}=0,
\label{eq:Qc-local}
\end{equation}
locates it to within $0.5\%$ at $l=2$, but the degeneracy is in fact exact. At
\begin{equation}
Q_c=\frac{2M\sqrt{L+1}}{L+2}\;\xrightarrow{\;l\gg1\;}\;\frac{2M}{l},
\label{eq:Qc}
\end{equation}
one has $s=M(5L+2)/(L+2)$, and the pseudoscalar potential coincides identically with the polar partner of the electromagnetic-led potential, $V_\Theta\equiv V_e^{(+)}$, where $V_e^{(\pm)}=\pm q_2\,dF/dr_*+q_2^2F^2+\mu^2(\mu^2+2)F$ and $F=\Delta/[r^3(\mu^2r+q_2)]$~\cite{Chandrasekhar1983}. By the Chandrasekhar isospectrality of $V_e^{(\pm)}$, $\omega_\Theta=\omega_e$ exactly at $Q_c$ ($Q_c/M=0.66144$, $0.51508$, $0.15861$ for $l=2,3,12$; numerically $|\omega_e-\omega_\Theta|<10^{-10}$). Below $Q_c$, $V_e<V_\Theta$ near the peak and the electromagnetic-led mode is lowered in frequency and more damped. Above $Q_c$ the shift reverses: $\mathrm{Re}\,\delta\omega_e>0$ and $\mathrm{Im}\,\delta\omega_e>0$, so this mode becomes \emph{longer-lived} [Fig.~\ref{fig:branches}(c)].

Near $Q_c$ the decoupled separation is linear, $\Delta_0\equiv\omega_e-\omega_\Theta\simeq(Q-Q_c)\,\Delta'$ with $M^2\Delta'=0.0961-0.0052\,i$ ($l=2$), and the coupled pair obeys the two-level form $\omega_\pm=\bar\omega\pm\tfrac12\sqrt{\Delta_0^2+(2g)^2}$. At $Q=Q_c$ the splitting is purely coupling-induced: $M|\omega_+-\omega_-|=0.4800/\sqrt\beta=2.41\sqrt\zeta$ ($l=2$) and $0.5762/\sqrt\beta=2.89\sqrt\zeta$ ($l=3$), constant to five digits for $10^3\le\beta\le10^{10}$ (units $M=\alpha=1$; Fig.~\ref{fig:avoided}). Isospectrality breaking is therefore $\mathcal{O}(\sqrt\zeta)$ at $Q_c$ for any coupling; the linear $\mathcal{O}(\zeta)$ shift of Eq.~\eqref{eq:shift-pert} applies only for $|Q-Q_c|\gg2g/|\Delta'|\simeq25\sqrt\zeta\,M$ ($l=2$). Across this window the two modes exchange character (Fig.~\ref{fig:avoided}b).

\section{Low-frequency limit and the static sector}
\label{sec:lowfreq}

The opposite regime, $\omega\to0$, provides an internal check: in the Regge--Wheeler channel it must reproduce the absence of a static effect found in the stationary analysis~\cite{RoutAltschulRN2026}, while isolating the dynamical origin of the present coupling.

\subsection{The source is linear in frequency}
\label{sec:lowfreq-source}

The normalisation $\Psi=i\Psi_{\mathrm{RW}}/\omega$ used in Sec.~\ref{sec:CdCS} is singular at
$\omega=0$; to probe the static limit one returns to the physical
Regge--Wheeler variable $\Psi_{\mathrm{RW}}$, in which the gravitational equation reads
\begin{equation}
\begin{aligned}
    &\left[\frac{d^2}{dr_*^2}+\omega^2-V_{\rm RW}\right]\Psi_{\mathrm{RW}} = S_{{\mathrm{RW}}} ,
  \\
  S_{\mathrm{RW}} &= -\,i\omega\,C_{\rm dCS}\,\Theta
      = -\,i\omega\,\frac{96\pi(Mr-Q^2)\Delta}{r^8}\,\Theta .
\end{aligned}
\label{eq:SQ}
\end{equation}
The source is \emph{exactly linear in $\omega$}. Through the electromagnetic channel in Eq.~\eqref{eq:CeTheta}, the Maxwell equation in physical variables also acquires a $\Theta$ source; the present statement concerns only the Regge--Wheeler channel. This is the physical content
concealed by the $\Psi=i\Psi_{\mathrm{RW}}/\omega$ redefinition, which absorbs the frequency into
the field and leaves $C_{\rm dCS}$ apparently $\omega$-independent. The factor
$\omega$ originates in the imaginary, $\omega$-proportional Cotton component
$C^{r\phi}\propto i\omega R_l$ of Sec.~\ref{sec:cotton}: the master-variable
reduction combines $C^{t\phi}$ and $C^{r\phi}$ so that the source in $Q$ emerges
cleanly proportional to $\omega$. The $\theta\varphi$ equation gives $h_0=(if/\omega)(fh_1)'$, so $h_1=\mathcal{O}(\omega h_0)$ and $\Psi=i f h_1/(\omega r)$ remains finite as $\omega\to0$; the factor $\omega$ in Eq.~\eqref{eq:SQ} reflects the vanishing of $\Psi_{\rm RW}$, not a singular normalization.

\subsection{Static limit of the Regge--Wheeler channel}
\label{sec:lowfreq-static}

Two consequences follow from Eq.~\eqref{eq:SQ}:
\begin{itemize}
\item \emph{Regge--Wheeler channel.} Within the linear axial sector, and at the order retained here, the direct pseudoscalar source of the Regge--Wheeler equation vanishes for strictly stationary perturbations, $S_{\rm RW}\to0$ as $\omega\to0$. This is consistent with the stationary analysis of Ref.~\cite{RoutAltschulRN2026}, now expressed as a frequency scaling. The statement does \emph{not} extend to the static tidal response. The static axial sector is carried by $h_0$, which is sourced by the $\omega$-independent component $C_{t\phi}$; and $\delta(\pRR)$ is nonzero for static $h_0$ [Eq.~\eqref{eq:dstarRR}]. Static axial (magnetic-type) Love numbers are therefore generically modified at $\mathcal{O}(\zeta)$, as found for Schwarzschild in Ref.~\cite{Cardoso2017TLN}; we do not compute them here.
\item \emph{Dynamical perturbations.} For $\omega\neq0$ the coupling is nonzero and $\mathcal{O}(\omega)$ in these variables. The quasinormal-mode effect is intrinsically dynamical.
\end{itemize}

\subsection{Order counting}
\label{sec:lowfreq-order}
Eliminating $\Theta$, the two couplings combine as $(-i\omega C_{\rm dCS})\,G_\Theta\,(\tfrac{i}{\omega}C_{\Theta g})=C_{\rm dCS}G_\Theta C_{\Theta g}$ with $G_\Theta=(\square+\omega^2-V_\Theta)^{-1}$: the factors of $\omega$ cancel, so the static $\mathcal{O}(\alpha)$ source vanishes while the quasinormal shift is $\mathcal{O}(\zeta)$ off resonance. The low-frequency limit fixes this scaling but does not by itself give a quasinormal shift, since the fundamental mode lies at $\omega\sim M^{-1}$.

\section{Near-extremal limit}
\label{sec:extremal}

The opposite corner of parameter space is the near-extremal black hole,
$Q\to M$, where the two horizons coalesce ($r_+\to r_-\to M$), the temperature
$T_H\to0$, and the near-horizon geometry approaches AdS$_2\times S^2$. We show
that, in contrast to the eikonal enhancement of Sec.~\ref{sec:eikonal}, the Chern--Simons effect on the near-horizon (zero-damping) family is \emph{suppressed} and vanishes at extremality, while photon-sphere modes are not suppressed.

\begin{figure*}[htbp]
\centering
\includegraphics[width=0.85\textwidth]{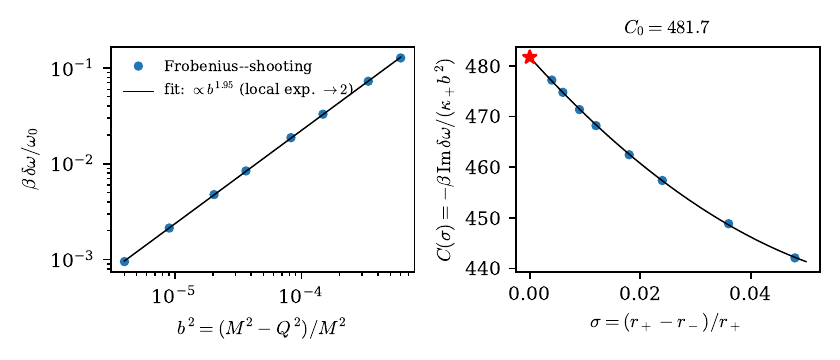}
\caption{Fundamental gravitational-led ZDM, $l=2$ (units $M=\alpha=1$). Left: $\beta\,\delta\omega/\omega_0$ versus $b^2=1-Q^2/M^2$ with a free power-law fit. Right: $C(\sigma)$ and its extrapolation $C_0=481.7$ (star).}
\label{fig:zdm}
\end{figure*}

\subsection{The source develops a node at the degenerate horizon}
\label{sec:extremal-node}

Writing $b\equiv\sqrt{M^2-Q^2}$, so that $r_\pm=M\pm b$ and extremality
is $b\to0$, the coupling combination $(Mr-Q^2)$ that controls \emph{both}
$C_{\rm dCS}$ and $C_{\Theta g}$ evaluated at the outer horizon is
\begin{equation}
  \big(Mr-Q^2\big)\big|_{r=r_+} = b\,r_+ = \sqrt{M^2-Q^2}\,r_+ ,
  \label{eq:node}
\end{equation}
which vanishes linearly in $b$. The Cotton source therefore acquires a node at the degenerate horizon --- precisely where the near-horizon (zero-damping) modes are
supported.

\subsection{Near-horizon scalings and conformal weights}
\label{sec:extremal-nh}

Introducing the near-horizon coordinate $x$ via $r-r_+=bx$, one has
$\Delta=b^2 x(x+2)$ and $(Mr-Q^2)\simeq bM(1+x)$, so both couplings are cubically suppressed (one power of $b$ beyond the $b^2$ carried by the potentials),
\begin{equation}
  C_{\rm dCS}\simeq\frac{96\pi\,x(x+1)(x+2)}{M^7}\,b^3,
  \quad C_{\Theta g}\propto b^3 .
  \label{eq:nh-couplings}
\end{equation}
Each field acquires an AdS$_2$ conformal weight $h$, fixed by the near-horizon value of its potential through $h(h-1)=U(r_+)/r_+$, where $V=(\Delta/r^5)U$. At extremality $(9M^2+4M^2\mu^2)=M^2(2l+1)^2$, so $q_1=2M(l+2)$ and $q_2=-2M(l-1)$, and
\begin{equation}
h_g=l,\qquad h_\Theta=l+1,\qquad h_e=l+2 .
\label{eq:weights}
\end{equation}
The corresponding zero-damping modes (ZDMs) are
\begin{equation}
\omega\simeq-i\kappa_+\,(h+n),\quad \kappa_+=\frac{r_+-r_-}{2r_+^2},
\label{eq:zdm}
\end{equation}
in agreement with Zimmerman and Mark~\cite{ZimmermanMark2016}, who established the existence of gravito-electromagnetic ZDMs of near-extremal RN. In their notation $|\delta_j|+\tfrac12=h$. These modes are purely damped and their decay rate vanishes at extremality. The integer spacing of Eq.~\eqref{eq:weights} implies near-degeneracies: the gravitational-led overtone $n=1$ ($h+n=l+1$) is degenerate with the pseudoscalar fundamental at extremality, and the electromagnetic-led ZDMs with pseudoscalar overtones. The \emph{fundamental} gravitational-led ZDM ($h+n=l$) is non-degenerate.

\subsection{The shift of the zero-damping modes}
\label{sec:extremal-shift}

In the throat both couplings carry one power of $b$ relative to the potentials [Eq.~\eqref{eq:nh-couplings}]. For a non-degenerate ZDM, second-order perturbation theory therefore gives $\delta\omega\sim\kappa_+\zeta\,b^2$, i.e.\ $\delta\omega/\omega\sim\zeta\,(M^2-Q^2)/M^2$.

This mode is isolated, so the branch-reduced system~\eqref{eq:reduced} applies. We have computed its shift with that system, using a Frobenius--shooting solver (Appendix~\ref{app:numerics}) validated against the ZDM frequencies of Ref.~\cite{ZimmermanMark2016}. For the fundamental gravitational-led ZDM with $l=2$:
\begin{itemize}
\item The shift is purely imaginary, $\beta\,\delta\omega=-i\,C\,\kappa_+b^2$.
\item $C=477.2,\,474.8,\,471.4,\,468.2,\,462.5,\,457.4,\,448.9,\\442.1$ for $\sigma\equiv(r_+-r_-)/r_+=0.004$--$0.048$, extrapolating to $C_0=481.7\pm0.6$ (fits over different $\sigma$ windows; units $M=\alpha=1$).
\item A fit with free exponent, $\delta\omega/\omega_0\propto b^{p}$, gives a local exponent $p=1.98$ at $\sigma=0.005$, extrapolating to $p=2.00$ as $\sigma\to0$ (Fig.~\ref{fig:zdm}); the small deviation is the $\mathcal{O}(\sigma)$ variation of $C(\sigma)$. Only purely damped roots ($\mathrm{Re}\,\omega=0$) that are continuous in $\sigma$, with $\omega/\kappa_+\to2$, are accepted.
\end{itemize}
Hence
\begin{equation}
\frac{\delta\omega}{\omega_0}=4\pi C_0\,\zeta\Big(1-\frac{Q^2}{M^2}\Big)\simeq6.0\times10^{3}\,\zeta\Big(1-\frac{Q^2}{M^2}\Big),
\label{eq:extremal-result}
\end{equation}
real and positive: the mode becomes \emph{more} damped, and the breaking vanishes at extremality. The large coefficient restricts Eq.~\eqref{eq:extremal-result} to $\zeta\ll2\times10^{-4}$.

For the near-degenerate overtones the suppression is weaker. The gravitational-led $n=1$ ZDM lies within $\Delta\omega/\kappa_+\simeq2\times10^{-4}$--$3\times10^{-3}$ of the pseudoscalar fundamental (for $\sigma=0.012$--$0.048$, scaling as $\sigma^2$). Its first-order shift, $\beta\,\delta\omega\simeq+(33\text{--}38)\,i\,\kappa_+$, is not suppressed by $b^2$. For realistic couplings the pair is in the avoided-crossing regime, where $\delta\omega/\omega=\mathcal{O}\big(\sqrt{\zeta(M^2-Q^2)}/M\big)$: still vanishing at extremality, but only as a square root.

\subsection{The two families together}
\label{sec:extremal-summary}

The node of the Cotton source at the degenerate horizon suppresses only the modes that live there. Photon-sphere modes remain relevant to the oscillatory ringdown; their relative excitation compared with ZDMs is not computed here. These modes are \emph{not} suppressed: their shift stays finite and grows as $Q\to M$ ($\beta\,\delta\omega=-0.97-0.35\,i$ at $Q/M=0.98$, $l=2$; Table~\ref{tab:shift}). Table~\ref{tab:limits} summarizes the limits.

\begin{table*}[htbp]
\centering
\caption{Limiting behavior of the dCS isospectrality-breaking shift.}
\label{tab:limits}
\begin{tabular}{lll}
\hline\hline
Regime & Mode family & Shift \\
\hline
Eikonal, off resonance, $Q=0$ & photon sphere (grav.-led) & $M\delta\omega\simeq-2.25\,\zeta\,l^3$ \\
Eikonal, off resonance, $Q>0$, $Ql\gg M$ & photon sphere (grav.-led) & $\propto\zeta\,l^2/Q$ \\
Strong mixing ($\zeta l^4\gtrsim1$ or $\zeta l^2\gtrsim Q^2/M^2$) & photon sphere & $\propto l\sqrt\zeta$ \\
$Q\simeq Q_c(l)$ & photon sphere (EM-led) & $\mathcal{O}(\sqrt\zeta)$, sign reversal across $Q_c$ \\
Near-extremal & photon sphere & not suppressed \\
Near-extremal & fundamental grav.-led ZDM ($l=2$) & $\delta\omega/\omega\simeq6.0\times10^3\,\zeta(1-Q^2/M^2)$ \\
Near-extremal & near-degenerate ZDM overtones & $\mathcal{O}\big(\sqrt{\zeta(M^2-Q^2)}/M\big)$ \\
\hline\hline
\end{tabular}
\end{table*}

\section{Weak-coupling limit and validity of the perturbative expansion}
\label{sec:weakcoupling}

The results of the preceding sections rest on treating the Chern--Simons
coupling perturbatively. We now make this expansion explicit, identify its
dimensionless control parameter, and delimit its regime of validity.

\subsection{The expansion parameter}
\label{sec:weak-param}

Restoring the coupling $\alpha$, the two off-diagonal couplings scale as
$C_{\rm dCS}\propto\alpha$ and $C_{\Theta g}\propto\alpha/\beta$, so every mode
shift is controlled by the product $C_{\rm dCS}C_{\Theta g}\propto\alpha^2/\beta$.
The natural expansion parameter of branch $a$ is the ratio of the weighted coupling to the squared level spacing,
\begin{equation}
\begin{aligned}
    \varepsilon_a&\equiv\frac{w_a^2\,C_{\rm dCS}C_{\Theta g}}{D_a^2}\\
    &=\frac{2304\pi\,\alpha^2\,(l-1)l(l+1)(l+2)\,(Mr-Q^2)^2\,w_a^2}{\beta\,r^{6}\,\big[2M-6Q^2/r+q_{(a)}\big]^2}.
\end{aligned}
\label{eq:eps}
\end{equation}
The horizon factor cancels, so $\varepsilon_a$ is smooth in $r$. For the gravitational-led branch, $\varepsilon_g\sim\zeta l^4$ at $Q=0$ and $\varepsilon_g\sim\zeta l^2(M/Q)^2$ at fixed $Q>0$ with $Ql\gg M$. For the electromagnetic-led branch, the local parameter $\varepsilon_e(r)$ becomes singular wherever the local potential spacing 
$D_e(r)$ vanishes. The exact spectral resonance occurs at Eq.~\eqref{eq:Qc}, where the scalar potential coincides with the polar partner of the axial electromagnetic-led potential. Near this resonance, perturbative validity must be assessed using the complex mode separation.

\subsection{Regime of validity}
\label{sec:weak-validity}

The $\mathcal{O}(\zeta)$ treatment of branch $a$ is controlled by $\varepsilon_a\ll1$. For the gravitational-led branch this requires
\begin{equation}
\zeta\ll1\quad\text{and}\quad
\begin{cases} l\ll\zeta^{-1/4}, & Q=0,\\ l\ll (Q/M)\,\zeta^{-1/2}, & Q>0,\ Ql\gg M.\end{cases}
\label{eq:validity}
\end{equation}
The perturbative treatment also fails near genuine degeneracies with pseudoscalar modes, where the shift becomes $\mathcal{O}(\sqrt\zeta)$:
\begin{itemize}
\item the electromagnetic-led branch at $Q\simeq Q_c(l)$ (Sec.~\ref{sec:eikonal-em});
\item the near-extremal ZDM overtones (Sec.~\ref{sec:extremal-shift}).
\end{itemize}
In practice the first-order shift of branch $a$ is reliable when it is small compared with the separation from the nearest pseudoscalar mode, $|\delta\omega_a|\ll|\omega_a-\omega_\Theta|$. Near $Q_c$ this means $|Q-Q_c|\gg25\sqrt\zeta\,M$ ($l=2$, Sec.~\ref{sec:eikonal-em}). For the fundamental ZDM the large coefficient in Eq.~\eqref{eq:extremal-result} restricts it to $\zeta\ll2\times10^{-4}$, a much narrower range than $\zeta\ll1$.

\subsection{Self-consistency}
\label{sec:weak-selfconsistency}
   The half-integer power $\zeta^{1/2}$ arises only where the pseudoscalar Green's function $G_\Theta$ becomes resonant: beyond the window~\eqref{eq:validity} at large $l$, and at the degeneracies listed above. Away from them the shift is the $\mathcal{O}(\zeta)$ quadrature
\begin{equation}
  \delta\omega_a = \frac{1}{2\omega_a}
     \frac{\displaystyle\int dr_*\;Z_a^{(0)}\,w_aC_{\rm dCS}\,G_\Theta\,w_aC_{\Theta g}\,Z_a^{(0)}}
          {\displaystyle\int dr_*\,\big(Z_a^{(0)}\big)^2},
  \label{eq:mcmanus}
\end{equation}
the coupled-equation form of the parametrized ringdown formalism~\cite{McManus2019Param}, with $Z_a^{(0)}$ the general-relativistic eigenfunction and the unconjugated (regularized) norm. Rather than regularizing the divergent quasinormal norm, we solve the coupled eigenproblem directly (Sec.~\ref{sec:wkb}).

\section{The unperturbed spectrum: baseline and validation}
\label{sec:numval}

Before evaluating the Chern--Simons shift we fix the unperturbed RN axial spectrum $\omega_0(Q)$, against which all corrections are measured. We compute it directly with the spectral method used for the coupled system (Appendix~\ref{app:numerics}), and we cross-check it against an established Schwarzschild solver, a perturbative charge expansion, and the continued-fraction results of Berti and Kokkotas~\cite{BertiKokkotas2005,Hatsuda2024,Stein2019}.

\subsection{Neutral anchor}
\label{sec:numval-anchor}

At $Q=0$ the axial gravitational sector reduces to the Regge--Wheeler equation,
whose fundamental $l=2$ quasinormal frequency is known to high precision. The
Leaver continued-fraction value,
\begin{equation}
\begin{aligned}
    \omega_0 M\big|_{Q=0} &= 0.3736716844 - 0.0889623157\,i\\
  &\text{for}\quad (l=2,\ n=0),
\end{aligned}
\label{eq:schw-anchor}
\end{equation}
is reproduced to all quoted digits by the \texttt{qnm} package (Kerr solver at
spin $a=0$)~\cite{Stein2019}, providing an independent check of the neutral endpoint.

In the Schwarzschild limit, $V_\Theta+\delta V_\Theta$ reproduces the potential $V_{22}$ of Ref.~\cite{Molina2010} [their Eq.~(3.6)], confirming the $\mathcal{O}(\alpha^2/\beta)$ back-reaction term.

\subsection{Charge dependence}
\label{sec:numval-charge}

For $Q>0$ we use the perturbative expansion of the fundamental frequency in the
squared charge $x\equiv(Q/M)^2$,
\begin{equation}
  \omega_0(Q)\,M = \sum_{k=0}^{\infty} w_k\,\Big(\frac{Q}{M}\Big)^{2k},
  \label{eq:pert-series}
\end{equation}
with coefficients (for $l=2$, $n=0$, axial gravitational)~\cite{Hatsuda2024}
\begin{equation}
  \begin{aligned}
    w_0 &= 0.3736716844 - 0.0889623157\,i, \\
    w_1 &= 0.0258176729 - 0.0028240321\,i, \\
    w_2 &= 0.0251877887 + 0.0002053245\,i, \\
    w_3 &= -0.0047481702 + 0.0025084021\,i, \\
    w_4 &= 0.0155726501 + 0.0004128797\,i.
  \end{aligned}
  \label{eq:pert-coeffs}
\end{equation}
The series~\eqref{eq:pert-series} is built on the Chandrasekhar-decoupled gravitational channel [Eq.~\eqref{eq:decoupled} with $q=q_1$], whose frequencies coincide with those of the coupled electromagnetic--gravitational system.

\subsection{Cross-validation}
\label{sec:numval-crosscheck}

Table~\ref{tab:numval} compares the series and the spectral solution with the continued-fraction results of Berti and Kokkotas~\cite{BertiKokkotas2005}. The spectral baseline agrees with the tabulated continued-fraction results within approximately $10^{-5}$ at all displayed charges.
\begin{table*}[htbp]
\centering
\begin{tabular}{ccccc}
\toprule
$Q/M$ & series & spectral (ECS) & Berti--Kokkotas & $|\Delta_{\rm ECS}|$\\
\midrule
$0.0$ & $0.37367-0.08896\,i$ & $0.37367-0.08896\,i$ & $0.37367-0.08896\,i$ & $<10^{-5}$\\
$0.2$ & $0.37474-0.08907\,i$ & $0.37474-0.08907\,i$ & $0.37474-0.08908\,i$ & $1\times10^{-5}$\\
$0.4$ & $0.37844-0.08940\,i$ & $0.37844-0.08940\,i$ & $0.37844-0.08940\,i$ & $<10^{-5}$\\
$0.6$ & $0.38627-0.08983\,i$ & $0.38622-0.08981\,i$ & $0.38622-0.08981\,i$ & $<10^{-5}$\\
$0.8$ & $0.40188-0.08996\,i$ & $0.40122-0.08964\,i$ & $0.40122-0.08964\,i$ & $<10^{-5}$\\
\bottomrule
\end{tabular}
\caption{Baseline RN axial gravitational-led fundamental ($l=2$, $n=0$), $\omega_0M$: charge expansion~\eqref{eq:pert-series}, the spectral (exterior-complex-scaling) solution, and the continued-fraction results of Berti and Kokkotas. The spectral solution agrees with Berti--Kokkotas at all charges. The series degrades beyond $Q/M\approx0.6$, near its radius of convergence.}
\label{tab:numval}
\end{table*}

\begin{figure*}[htbp]
\centering
\includegraphics[width=0.75\textwidth]{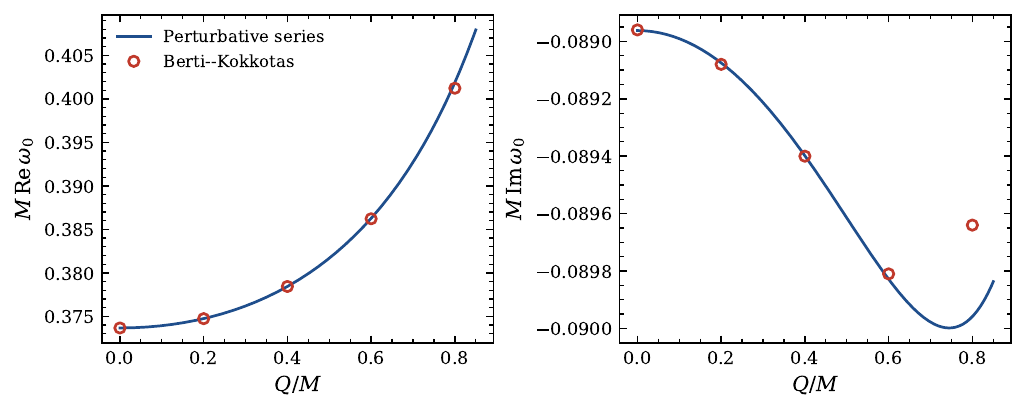}
\caption{Baseline Reissner--Nordstr\"om axial gravitational fundamental
($l=2$, $n=0$): real (left) and imaginary (right) parts of $M\omega_0$ versus
charge. The solid curve is the perturbative charge expansion
[Eq.~\eqref{eq:pert-series}]; open circles are the independent
continued-fraction results of Berti and Kokkotas~\cite{BertiKokkotas2005}.
The truncated series agrees with the continued-fraction data to five decimal places
for $Q/M\lesssim0.4$ and degrades towards its radius of convergence. The
exterior-complex-scaling baseline used in this work agrees with the
continued-fraction values at all displayed charges (Table~\ref{tab:numval}).}
\label{fig:baseline}
\end{figure*}

As shown in Fig.~\ref{fig:baseline}, the expansion~\eqref{eq:pert-series} loses accuracy for $Q/M\gtrsim0.7$, consistent with its estimated radius of convergence $x\approx0.64$~\cite{Hatsuda2024}. All shifts below are therefore computed on the spectral baseline. Its agreement with Ref.~\cite{BertiKokkotas2005} holds at all charges, and with the series to $8\times10^{-9}$ for $Q/M\le0.1$.

\section{Frequency shifts of the coupled system}
\label{sec:wkb}

\subsection{Method}
\label{sec:wkb-method}

In the normalization $\Psi=i\Psi_{\rm RW}/\omega$ all couplings in Eq.~\eqref{eq:coupled} are algebraic and independent of $\omega$. The coupled problem is therefore a linear eigenvalue problem in $\omega^2$. We solve it by exterior complex scaling~\cite{AguilarCombes1971,BalslevCombes1971,Simon1979,Moiseyev1998}:
\begin{itemize}
\item the tortoise coordinate is continued to $r_*=e^{i\theta}s$, which turns the quasinormal boundary conditions into decay at both ends;
\item the operator is discretized by Chebyshev collocation in $s$.
\end{itemize}
The linear coefficient $\beta\,\delta\omega$ is extracted from the difference of coupled and uncoupled eigenvalues on the same grid, with Richardson extrapolation in $1/\beta$. The method, its gates, and convergence tests are given in Appendix~\ref{app:numerics}. The eigenvalue calculation retains the full radial coupled system and does not invoke the local Green-function approximation used only in the asymptotic discussion of Sec.~\ref{sec:eikonal} and Appendix~\ref{app:wkb}. The discretization error is negligible: the coefficients change by at most $4\times10^{-5}$ (relative) under changes of the contour angle, box size, resolution and the pivot of the complex rotation ($\le7\times10^{-6}$ for $\theta\ge0.6$, $S\ge60$), and an independent shooting integration of the full three-field system reproduces them to $1.4\times10^{-6}$. The dominant uncertainty is the extrapolation in $1/\beta$, bounded by $4\times10^{-4}$ for $l=2$ (Table~\ref{tab:shift}) and by $0.4\%$ for the $l=6$ coefficients used in the scaling analysis (Table~\ref{tab:convergence}).

\subsection{Neutral limit}
\label{sec:wkb-neutral}

For $l=2$, $n=0$,
\begin{equation}
\beta\,\delta\omega\big|_{Q=0}=-0.7733-0.3946\,i\qquad(M=\alpha=1),
\label{eq:q0-result}
\end{equation}
i.e.\ $M\delta\omega=(-19.4-9.9\,i)\,\zeta$. This agrees with our independent evaluation of the frequency-domain formula of Molina \emph{et al.}~\cite{Molina2010}, $-0.7731-0.3945\,i$, which reproduces their numerical shift at $M^4\beta=100$ to $0.4\%$. The oscillation frequency is lowered, and the axial mode is \emph{more} strongly damped than its polar partner. For higher multipoles at $Q=0$:
\begin{center}
\begin{tabular}{ccc}
\toprule
$l$ & $\beta\,\mathrm{Re}\,\delta\omega$ & $\beta\,\mathrm{Im}\,\delta\omega$\\
\midrule
2 & $-0.7733$ & $-0.3946$\\
3 & $-2.872$ & $-0.519$\\
4 & $-6.858$ & $-0.653$\\
5 & $-13.27$ & $-0.809$\\
6 & $-22.62$ & $-0.991$\\
\bottomrule
\end{tabular}
\end{center}
These approach the $l^3$ law~\eqref{eq:domega-schw} with $\mathcal{O}(1/l)$ corrections (ratio $1.17$ at $l=6$, $1.10$ at $l=12$). A local-WKB proxy (Appendix~\ref{app:wkb}) gives $\beta\,\delta\omega\simeq-0.45+0.15\,i$ at $Q=0$: the correct sign of the real part but the wrong sign of the imaginary part, and it misses the non-monotonic charge dependence; we use it only as a qualitative check.

\subsection{Charge dependence}
\label{sec:wkb-result}

\begin{figure*}[t]
\centering
\includegraphics[width=0.85\textwidth]{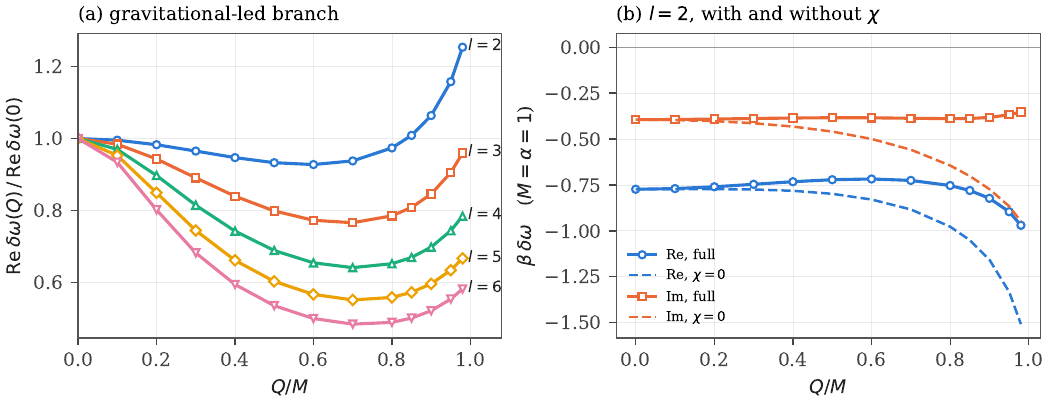}
\caption{Isospectrality-breaking shift of the gravitational-led fundamental from the coupled spectral solution (units $M=\alpha=1$).
(a)~$\mathrm{Re}\,\delta\omega(Q)/\mathrm{Re}\,\delta\omega(0)$ for $l=2$--$6$.
(b)~$l=2$: real and imaginary parts with (solid) and without (dashed) the electromagnetic channel $\chi$.}
\label{fig:shift}
\end{figure*}

Table~\ref{tab:shift} and Fig.~\ref{fig:shift} give the shift of the gravitational-led fundamental across the full charge range. Three features stand out.
\begin{itemize}
\item \emph{$l=2$ is non-monotonic.} $|\mathrm{Re}\,\delta\omega|$ decreases by $7\%$ to a minimum near $Q/M\simeq0.6$, then grows to $1.25$ times its neutral value at $Q/M=0.98$.
\item \emph{Higher multipoles are strongly suppressed at intermediate charge.} The shift falls to $0.49$ of its neutral value at $l=6$ near $Q/M\simeq0.7$. This is the $l^3\to l^2/Q$ change of scaling of Sec.~\ref{sec:eikonal}.
\item \emph{The electromagnetic channel matters quantitatively.} Without it ($\chi=0$), the $l=2$ shift would grow monotonically to $1.95$ times its neutral value, and $|\mathrm{Im}\,\delta\omega|$ by a factor $2.4$. With it, $\mathrm{Im}\,\delta\omega$ stays nearly constant.
\end{itemize}
In all cases $\mathrm{Im}\,\delta\omega<0$, so charge does not change the sign of the damping correction of the gravitational-led modes.

\begin{table*}[htbp]
\centering
\begin{tabular}{cccc}
\toprule
$Q/M$ & $\omega_0M$ & $\beta\,\delta\omega$ & $\beta\,\delta\omega\,(\chi=0)$\\
\midrule
$0.0$ & $0.37367-0.08896\,i$ & $-0.7733-0.3946\,i$ & $-0.7733-0.3946\,i$\\
$0.2$ & $0.37474-0.08907\,i$ & $-0.7599-0.3913\,i$ & $-0.7726-0.4026\,i$\\
$0.4$ & $0.37844-0.08940\,i$ & $-0.7323-0.3854\,i$ & $-0.7817-0.4320\,i$\\
$0.6$ & $0.38622-0.08981\,i$ & $-0.7173-0.3845\,i$ & $-0.8291-0.4995\,i$\\
$0.8$ & $0.40122-0.08964\,i$ & $-0.7531-0.3891\,i$ & $-0.9769-0.6439\,i$\\
$0.9$ & $0.41357-0.08833\,i$ & $-0.8225-0.3825\,i$ & $-1.1571-0.7731\,i$\\
$0.98$ & $0.42731-0.08498\,i$ & $-0.9694-0.3529\,i$ & $-1.5075-0.9477\,i$\\
\bottomrule
\end{tabular}
\caption{Isospectrality-breaking shift of the gravitational-led fundamental ($l=2$, $n=0$) from the coupled spectral solution (units $M=\alpha=1$; physical shift $=$ tabulated value$/\beta$). The last column omits the electromagnetic channel $\chi$ of Eq.~\eqref{eq:coupled}; it is a diagnostic truncation, not the finite-charge dCS system derived in this work.}
\label{tab:shift}
\end{table*}

\section{Conclusions}
\label{sec:conclusions}

We have derived the axial couplings of a dynamical Chern--Simons pseudoscalar to a charged black hole and computed the resulting breaking of axial--polar isospectrality in the Reissner--Nordstr\"om quasinormal spectrum. Beyond the charge-dependent kernel, three features have no Schwarzschild analog: the shifted electromagnetic-led branch and its exact degeneracy with the pseudoscalar at $Q_c$, the control that charge exerts on how the coupling is shared between the two branches, and the opposite near-extremal behavior of zero-damping and photon-sphere modes.

\emph{Couplings.} By direct reduction of the linearized field equations we obtained, in closed form, the coupling $C_{\rm dCS}=96\pi(Mr-Q^2)\Delta/r^8$ of the pseudoscalar to the axial gravitational field and its reciprocal $C_{\Theta g}$. Both are built from the kernel $(Mr-Q^2)\Delta/r^8$, which reflects the single nonvanishing Weyl scalar $\Psi_2$ of the background. At finite charge the pseudoscalar couples, in both directions, to the combination $\Psi+\chi\Psi_e$, with $\chi=2Q/[r\sqrt{(l-1)(l+2)}]$. This electromagnetic channel is absent for Schwarzschild, yet changes the charge dependence of the shift qualitatively. Because the Einstein--Maxwell sector decouples under a constant rotation, the leading shift of each RN branch is governed exactly by a single weighted coupling.

\emph{Shifts.} We computed the shifts with an exterior-complex-scaling spectral method and, for near-extremal zero-damping modes, a Frobenius--shooting method validated against Ref.~\cite{ZimmermanMark2016}. In the neutral limit $\beta\,\delta\omega=-0.7733-0.3946\,i$ ($l=2$, $M=\alpha=1$), in agreement with Ref.~\cite{Molina2010}. Charge changes the off-resonance scaling of the gravitational-led shift from $l^3$ to $l^2/Q$ and makes it non-monotonic; it drives the electromagnetic-led branch through an exact degeneracy with the pseudoscalar at $Q_c=2M\sqrt{L+1}/(L+2)$, where the breaking is $\mathcal{O}(\sqrt\zeta)$ ($M|\Delta\omega|\simeq2.4\sqrt\zeta$ for $l=2$) and reverses sign; and near extremality it suppresses the fundamental zero-damping mode, $\delta\omega/\omega_0\simeq6.0\times10^{3}\zeta(1-Q^2/M^2)$ for $l=2$, while the photon-sphere shifts stay finite and the near-degenerate overtones scale only as $\sqrt{\zeta(M^2-Q^2)}/M$. The gravitational-led photon-sphere modes and the fundamental zero-damping mode become more damped, the zero-damping overtone less damped. The direct source of the Regge--Wheeler channel vanishes linearly as $\omega\to0$; the static tidal response through $h_0$ is not addressed here.

\emph{Scope.} All shifts are at leading order in $\zeta$ off resonance (or at leading order in $\sqrt\zeta$ near degeneracies), and the near-degenerate ZDM overtones are characterized only by their scaling. Table~\ref{tab:status} summarizes the status of each result.

\begin{table*}[htbp]
\centering
\caption{Status of results.}
\label{tab:status}
\begin{tabular}{lll}
\hline\hline
Result & Method & Status\\
\hline
Polar sector & parity; $\delta(\pRR)|_{\rm polar}=0$ & exact (linear order)\\
$C_{\rm dCS}$ at finite $Q$ & direct reduction; $t\phi$ consistency & exact\\
\cell{$C_{\Theta g}$ and the $\chi$ channel\\ at finite $Q$} & \cell{direct reduction of $\delta(\pRR)$\\ and Einstein--Maxwell} & exact\\
Branch decoupling and weights $w_{g,e}$ & $r$-independent mixing matrix & exact\\
Branch-reduced system~\eqref{eq:reduced} & \cell{branch projection;\\ numerical comparison ($10^{-7}$)} & \cell{equivalent for isolated branch\\ shifts through $\mathcal{O}(\zeta)$;\\ verified away from degeneracies}\\
\cell{$\delta\omega(Q)$, photon-sphere modes,\\ $l=2$--$6$} & spectral (ECS) eigensolve & \cell{numerical; $\le4\times10^{-4}$ ($l=2$),\\ $\le0.4\%$ ($l=6$)}\\
Neutral $l=2$ shift & ECS; Molina formula & numerical; agrees with Ref.~\cite{Molina2010}\\
Scalings $l^3$ ($Q=0$), $l^2/Q$ ($Q>0$) & local branch formula & asymptotic; $\leq 1\%$ at $l=12$\\
Approach to $l^3$, $l^2/Q$ laws & reduced ECS to $l=30$ & numerical; $\le1.3\%$ after $1/l$ extrapolation\\
$Q_c(l)$ and EM-led sign reversal & \cell{$V_\Theta\equiv V_e^{(+)}$ (exact);\\ ECS, shooting} & exact\\
$\mathcal{O}(\sqrt\zeta)$ splitting at $Q_c$ & ECS, $\beta=10^3$--$10^{10}$ & numerical, five digits\\
Fundamental ZDM shift & \cell{Frobenius--shooting,\\ validated vs Ref.~\cite{ZimmermanMark2016}} & numerical\\
Near-degenerate ZDM overtones & scaling + first-order numerics & estimate\\
Regge--Wheeler source $\propto\omega$ & exact & exact (linear axial sector)\\
Local-WKB values (App.~\ref{app:wkb}) & local WKB + local $G_\Theta$ & qualitative proxy\\
\hline\hline
\end{tabular}
\end{table*}

\emph{Outlook.} A natural extension is the rotating Kerr--Newman case, where rotation induces parity-violating effects already at first order in spin and coexists with the charge-driven effects found here. Whereas the quasinormal modes of \emph{uncharged} rotating black holes in dCS gravity have been computed by slow-rotation and spectral methods~\cite{Wagle2022,Srivastava2021,Li2025,ChungLamYunes2025}, the charged rotating case remains open. As black-hole spectroscopy sharpens, isospectrality breaking offers a characteristic signature of the dCS coupling in the ringdown, distinct from a mere shift of individual frequencies.

\begin{acknowledgments}
I thank Brett Altschul for helpful comments on the manuscript.
In preparing this work I used Claude Opus~5.5 (Anthropic), accessed through the claude.ai interface, under my direction. It assisted with (i) writing and debugging the numerical codes (the exterior-complex-scaling Chebyshev solver and the Frobenius/complex-ray shooting solver) and the scripts that produce the tables and figures; and (ii) drafting and editing parts of the text.
\end{acknowledgments}

\section*{Data availability}
The data that support the findings of this article are openly available~\cite{RNdCS_code}. The archive (GitHub repository \url{https://github.com/AbhiRout-cosmo/RN-dCS}, release v1.1.0, archived on Zenodo with version DOI \href{https://doi.org/10.5281/zenodo.23331188}{10.5281/zenodo.23331188}; all versions: \href{https://doi.org/10.5281/zenodo.23283164}{10.5281/zenodo.23283164}) contains the following material, released under the MIT license:
\begin{itemize}
\item[(i)] the \textsc{SymPy} derivations of the linearized Pontryagin density, the axial Einstein--Maxwell equations, the $C$-tensor components, the reduction to the coupled $\{\Psi,\Psi_e,\Theta\}$ master system, and the on-shell consistency check of the $t\phi$ equation (Secs.~\ref{sec:cotton}--\ref{sec:CThetag}, Appendices~\ref{app:cotton}--\ref{app:reciprocal});
\item[(ii)] the exterior-complex-scaling Chebyshev solver, the branch-reduced $\{Z_a,\Theta\}$ system, and the Frobenius/complex-ray shooting solver for the zero-damping modes (Appendix~\ref{app:numerics});
\item[(iii)] one script per numerical result, together with the data behind all tables and Figs.~\ref{fig:branches}, \ref{fig:eikratio}, \ref{fig:avoided}, \ref{fig:zdm} and~\ref{fig:shift}.
\end{itemize}
Every numerical value quoted in this article can be regenerated from these scripts.

\appendix
\section{Axial Cotton tensor components and their symbolic verification}
\label{app:cotton}

This appendix records the linearized $C$-tensor components that source the axial gravitational sector [Eqs.~\eqref{eq:Ctphi_final}--\eqref{eq:Crphi_final}], and the two limits used as cross-checks. All tensor algebra was carried out with a computer-algebra implementation (\textsc{SymPy}), which is available with the numerical codes (Appendix~\ref{app:numerics}).

\subsection{Setup and conventions}
\label{app:cotton-setup}

We perturb the static Reissner--Nordstr\"om (RN) background
\begin{equation}
\begin{aligned}
    ds^2&=-f\,dt^2+f^{-1}dr^2+r^2\big(d\theta^2+\sin^2\theta\,d\varphi^2\big),\\
    f&=\frac{\Delta}{r^2},\quad \Delta=r^2-2Mr+Q^2,
\end{aligned}
\label{eq:appA-bg}
\end{equation}
by a \emph{dynamical} pseudoscalar
\begin{equation}
\delta\vartheta=R(r)\,Y_l(\theta)\,e^{-i\omega t},
\label{eq:appA-scalar}
\end{equation}
with vanishing background scalar (the RN background solves the dCS field
equations exactly, since $\pRR|_{\rm RN}=0$). The C-tensor is
\begin{equation}
C^{\mu\nu}=(\nabla_\rho\vartheta)\,
\varepsilon^{\rho\sigma\tau(\mu}\nabla_\tau R^{\nu)}{}_{\sigma}
+(\nabla_\rho\nabla_\sigma\vartheta)\,\dual{R}{}^{\sigma(\mu\nu)\rho},
\label{eq:appA-ctensor}
\end{equation}
with the round brackets denoting symmetrisation \emph{with weight $1/2$}; this
normalisation is what fixes the gravitational-sector coupling at $96\pi$ rather
than $192\pi$ (Sec.~\ref{sec:CdCS}). The Levi-Civita tensor is $\varepsilon_{\mu\nu\rho\sigma}=\sqrt{-g}\,[\mu\nu\rho\sigma]$ with $[tr\theta\varphi]=+1$ (so $\varepsilon^{tr\theta\varphi}=-1/\sqrt{-g}$).

The axial (odd-parity) sector is generated by the vector harmonic
\begin{equation}
S^\varphi=\frac{\partial_\theta Y_l}{\sin\theta},
\label{eq:appA-harmonic}
\end{equation}
i.e. the contravariant component of $X_A=(-\varepsilon_A{}^B\partial_B Y_l)$ with
$\varepsilon_{AB}$ the Levi-Civita tensor on the unit two-sphere. Every nonzero
Cotton component below is proportional to $S^\varphi$; retaining the factor
$1/\sin\theta$ is essential and is what makes the projection onto the
gravitational sector diagonal in $l$ (Sec.~\ref{sec:CdCS}).

\subsection{The linearised components}
\label{app:cotton-components}

Evaluating \eqref{eq:appA-ctensor} on \eqref{eq:appA-bg}--\eqref{eq:appA-scalar}
gives two nonzero contravariant components in the axial sector; the algebra
naturally separates into the two pieces of \eqref{eq:appA-ctensor}, which we label $C_{(1)}$ (from the $\nabla\vartheta\,\varepsilon\nabla R$ term, which involves only the Ricci tensor) and $C_{(2)}$ (from the $\nabla\nabla\vartheta\,\dual{R}$ term).

\paragraph{The $t\varphi$ component.}
The two pieces are
\begin{align}
C^{t\varphi}_{(2)}
&=-\frac{3\,(Mr-Q^2)}{r^6}\Big(R'-\frac{R}{r}\Big)\,S^\varphi,
\label{eq:appA-C2tphi}\\
C^{t\varphi}_{(1)}
&=-\frac{3\,Q^2}{r^7}\,R\,S^\varphi.
\label{eq:appA-C1tphi}
\end{align}
Adding,
\begin{equation}
C^{t\varphi}=\left[\,-\frac{3(Mr-Q^2)}{r^6}\,R'
+\frac{3(Mr-2Q^2)}{r^7}\,R\,\right]S^\varphi ,
\label{eq:appA-Ctphi}
\end{equation}
where the shift $Q^2\to2Q^2$ in the $R$ coefficient relative to the $R'$ coefficient is the contribution of the Ricci piece $C_{(1)}$.

\paragraph{The $r\varphi$ component.}
Here only the $C_{(2)}$ piece contributes, and it is proportional to $\omega$:
\begin{equation}
C^{r\varphi}=C^{r\varphi}_{(2)}
=-\frac{3\,i\omega\,(Mr-Q^2)}{r^6}\,R\;S^\varphi,
\quad C^{r\varphi}_{(1)}=0.
\label{eq:appA-Crphi}
\end{equation}
The factor $i\omega$ is the algebraic origin of the \emph{dynamical} character of
the effect: it vanishes for a stationary ($\omega=0$) perturbation, and it is the
channel whose $\omega$ is absorbed by the master-variable normalisation
$\Psi=i\Psi_{\rm RW}/\omega$ to render the coupling algebraic
(Sec.~\ref{sec:CdCS}; see also the low-frequency analysis,
Sec.~\ref{sec:lowfreq}).

\paragraph{The $\theta\varphi$ component.}
\begin{equation}
C^{\theta\varphi}=0 .
\label{eq:appA-Cthetaphi}
\end{equation}
The absence of a tensor-harmonic ($\theta\varphi$) source means the pseudoscalar
does not excite the higher angular structure; the axial coupling is carried
entirely by the vector-harmonic components
\eqref{eq:appA-Ctphi}--\eqref{eq:appA-Crphi}.

\subsection{Internal consistency and symbolic verification}
\label{app:cotton-verify}

Three internal checks, all confirmed symbolically, fix confidence in
\eqref{eq:appA-Ctphi}--\eqref{eq:appA-Cthetaphi}.

\emph{(i) Piecewise reconstruction.} The separate pieces
\eqref{eq:appA-C2tphi}--\eqref{eq:appA-C1tphi} reproduce the combined component
\eqref{eq:appA-Ctphi} term by term: the $R'$ coefficient is
$-3(Mr-Q^2)/r^6$ from $C_{(2)}$ alone, while the $R$ coefficient $3(Mr-2Q^2)/r^7$ arises as $3(Mr-Q^2)/r^7-3Q^2/r^7$, i.e.\ the $-R/r$ term of $C_{(2)}$ plus $C_{(1)}$.

\emph{(ii) Harmonic structure.} Every component carries the single angular factor
$S^\varphi=\partial_\theta Y_l/\sin\theta$ and no other; in particular there is no
$Y_l$ (scalar) or tensor-harmonic admixture. Because the background is
spherically symmetric, the map
$\delta\vartheta\mapsto C_{\mu\nu}[\delta\vartheta]$ is $SO(3)$-equivariant and
cannot mix multipoles, so the projection onto the gravitational sector is diagonal
in $l$ rather than banded.

\emph{(iii) Curvature-invariant sub-checks.} The full set of background Riemann
components entering \eqref{eq:appA-ctensor}, the dual $\dual{R}{}^{\mu\nu\rho\sigma}$,
and the covariant derivatives $\nabla\nabla\vartheta$ were computed and confirmed
symbolically; the same implementation was used to cross-check the Pontryagin
structure of the reciprocal (scalar-sector) source in Appendix~\ref{app:reciprocal}.

\subsection{Neutral and extremal limits}
\label{app:cotton-limits}

\paragraph{Schwarzschild limit ($Q\to0$).}
The components reduce to
\begin{equation}
\begin{aligned}
        C^{t\varphi}\big|_{Q=0}&=\left[-\frac{3M}{r^5}R'+\frac{3M}{r^6}R\right]S^\varphi,\\
    C^{r\varphi}\big|_{Q=0}&=-\frac{3i\omega M}{r^5}\,R\,S^\varphi,
\end{aligned}
\label{eq:appA-Q0}
\end{equation}
which are the source components that reduce, in the master-variable calculation
of Sec.~\ref{sec:CdCS}, to the Schwarzschild dCS coupling
$C_{\rm dCS}\to 96\pi M f/r^5$ of Molina \emph{et al.}~\cite{Molina2010}. This is
the component-level form of the calibration presented in the main text.

\paragraph{Extremal node.}
The combination $(Mr-Q^2)$ common to
\eqref{eq:appA-Ctphi}--\eqref{eq:appA-Crphi}, evaluated at the outer horizon
$r_+=M+b$ with $b\equiv\sqrt{M^2-Q^2}$ and $Q^2=M^2-b^2$, is
\begin{equation}
\big(Mr-Q^2\big)\big|_{r=r_+}=b\,(M+b)=b\,r_+ \;\xrightarrow{\;b\to0\;}\;0 .
\label{eq:appA-node}
\end{equation}
The Cotton source therefore develops a node at the degenerate horizon as
extremality is approached, which is the microscopic origin of the suppression of the zero-damping modes found in Sec.~\ref{sec:extremal}.

\section{Reduction to the master equation and the coupling \texorpdfstring{$C_{\rm dCS}$}{C\_dCS}}
\label{app:reduction}

This appendix carries out, in detail, the reduction of the axial gravitational
field equations to a single master equation, and extracts the dynamical
Chern--Simons coupling $C_{\rm dCS}$ quoted in Sec.~\ref{sec:CdCS}. The steps are:
(i) assemble the two first-order axial field equations with the Cotton source of
Appendix~\ref{app:cotton}; (ii) eliminate the non-dynamical metric function
$h_0$; (iii) pass to the Regge--Wheeler master variable and adopt the
normalisation $\Psi=i\Psi_{\rm RW}/\omega$ that renders the coupling algebraic;
(iv) use the scalar equation of motion on shell to remove the residual
$\Theta''$; and (v) read off $C_{\rm dCS}$. Every intermediate expression below
was confirmed symbolically.

\subsection{The axial field equations with source}
\label{app:red-fieldeqs}

In Regge--Wheeler gauge the odd-parity metric perturbation is carried by
$h_0(r)=h_{t\varphi}$ and $h_1(r)=h_{r\varphi}$, each multiplying the axial vector
harmonic $W\equiv\sin\theta\,\partial_\theta Y_l$ with time dependence
$e^{-i\omega t}$. The linearised Einstein tensor on the static RN background
\eqref{eq:appA-bg}, after use of the harmonic identity
$W''=\cot\theta\,W'-L\,W$ (with $L\equiv l(l+1)$; the residual $W'$ terms cancel
identically), gives

\begin{align}
\delta G_{t\varphi}
&= W\!\Big[-\frac{\Delta}{2r^2}h_0''
-\frac{i\omega\Delta}{2r^2}h_1'
+\frac{Lr^2-4Mr+4Q^2}{2r^4}h_0 \nonumber \\
&\quad -\frac{i\omega\Delta}{r^3}h_1\Big],
\label{eq:appB-dGtphi}\\[4pt]
\delta G_{r\varphi}
&= W\!\left[\frac{i\omega r^2}{2\Delta}h_0'
-\frac{i\omega r}{\Delta}h_0
+\frac{\mathcal{N}(r)}{2r^4\Delta}h_1\right].
\label{eq:appB-dGrphi}
\end{align}
with
\begin{equation}
\mathcal{N}(r)=L\,r^2\Delta+4Mr(r^2-Q^2)-2(r^2-Q^2)(r^2+Q^2)-\omega^2 r^6 .
\label{eq:appB-N}
\end{equation}
From Eq.~\eqref{eq:einstein}, the dCS contribution to the axial field equations is $\delta R_{ab}\supset-2\kappa\alpha\,C_{ab}=-16\pi\alpha\,C_{ab}$; since
the background is Ricci-scalar flat and the pseudoscalar stress-energy is
$O(\vartheta^2)$, the off-diagonal axial components obey $\delta G_{\mu\varphi}=-16\pi\alpha\,C_{\mu\varphi}+\kappa\,\delta T^{\rm EM}_{\mu\varphi}$, the last term producing $C_Q$. Setting $\alpha=1$ and using
the covariant Cotton components of Appendix~\ref{app:cotton} [the coefficients of
$W$ in $C_{\mu\varphi}=g_{\mu\mu}g_{\varphi\varphi}C^{\mu\varphi}$], the source
terms are

\begin{align}
-16\pi\,C_{t\varphi}\big|_W
&=-\frac{48\pi f}{r^5}\Big[r(Mr-Q^2)R'-(Mr-2Q^2)R\Big]\nonumber\\
&\equiv S_t^{\rm raw},
\label{eq:appB-Straw}\\
-16\pi\,C_{r\varphi}\big|_W
&=\frac{48\pi\,i\omega\,(Mr-Q^2)}{f\,r^4}\,R\equiv S_r^{\rm raw}.
\label{eq:appB-Srraw}
\end{align}
(These follow from lowering Eqs.~\eqref{eq:appA-Ctphi}--\eqref{eq:appA-Crphi} with $g_{tt}=-f$, $g_{rr}=f^{-1}$, $g_{\varphi\varphi}=r^2\sin^2\theta$. The coefficient $96\pi$ of $C_{\rm dCS}$ emerges after the reduction below.)

\subsection{Elimination of \texorpdfstring{$h_0$}{h0}}
\label{app:red-elim}

Dividing \eqref{eq:appB-dGtphi} by $-\Delta/(2r^2)$ and \eqref{eq:appB-dGrphi} by
$i\omega r^2/(2\Delta)$ puts the system in the normalised form
\begin{align}
&h_0''+i\omega\,h_1'+A\,h_0+B\,h_1=S_t,
\label{eq:appB-I}\\
&h_0'=\frac{2}{r}h_0+C\,h_1+S_r,
\label{eq:appB-II}
\end{align}
with
\begin{equation}
A=-\frac{Lr^2-4Mr+4Q^2}{r^2\Delta},\quad
B=\frac{2i\omega}{r},\quad
C=\frac{i\,\mathcal{N}}{\omega r^6},
\label{eq:appB-ABC}
\end{equation}
and the normalised sources
\begin{equation}
S_t=-\frac{2r^2}{\Delta}\,S_t^{\rm raw},\qquad
S_r=-\frac{2i\Delta}{\omega r^2}\,S_r^{\rm raw}.
\label{eq:appB-normsrc}
\end{equation}
Differentiating \eqref{eq:appB-II} and substituting it and \eqref{eq:appB-II}
into \eqref{eq:appB-I} eliminates $h_0''$ and $h_0'$, leaving $h_0$ algebraically
determined,
\begin{equation}
h_0=\frac{S_t-S_r'-\tfrac{2}{r}S_r-(C+i\omega)h_1'
-\big(\tfrac{2C}{r}+C'+B\big)h_1}{\,2/r^2+A\,},
\label{eq:appB-h0}
\end{equation}
where $2/r^2+A=\big[(2-L)r^2-2Q^2\big]/(r^2\Delta)$. Re-inserting
\eqref{eq:appB-h0} into \eqref{eq:appB-II} yields a single second-order equation
for $h_1$; its homogeneous part is the Regge--Wheeler operator, and its
inhomogeneous part is the reduced source. Writing the homogeneous piece in terms
of the master variable
\begin{equation}
\Psi_{\rm RW}\equiv\frac{\Delta}{r^3}\,h_1,
\label{eq:appB-master}
\end{equation}
one recovers the Schr\"odinger type equations
$[\,d^2/dr_*^2+\omega^2-V_{\rm RW}\,]\Psi_{\rm RW}=\mathcal{S}$; the neutral limit
of this reduction was checked to reproduce the Schwarzschild Regge--Wheeler
potential $f[l(l+1)/r^2-6M/r^3]$ to all orders, fixing the elimination
combination independently of any assumed coefficient.

\subsection{Algebraic normalisation and the reduced source}
\label{app:red-normalisation}

The reduced source $\mathcal{S}$ contains both $\Theta$ and $\Theta'$ and depends
on $\omega$ through the $C^{r\varphi}\propto i\omega$ channel of
\eqref{eq:appB-Srraw}. Adopting the master function of Sec.~\ref{sec:CdCS},
\begin{equation}
\Psi\equiv\frac{i\,\Psi_{\rm RW}}{\omega},
\label{eq:appB-Psi}
\end{equation}
is not cosmetic: the factor $i/\omega$ absorbs the explicit $\omega$ of the
$r\varphi$ channel, so that the source of the $\Psi$-equation becomes independent
of $\omega$ and \emph{algebraic} in $\Theta$. Concretely, if $S_{\Psi_{\rm RW}}$
denotes the source of the $\Psi_{\rm RW}$-equation, then the source of the
$\Psi$-equation is $(i/\omega)\,S_{\Psi_{\rm RW}}$, and the two factors of
$\omega$ (one from $S_r^{\rm raw}$, one from the normalisation) cancel. This is
also the mechanism behind the exact frequency-linearity of the source in the
low-frequency analysis of Sec.~\ref{sec:lowfreq}.

\subsection{On-shell reduction and the coupling}
\label{app:red-onshell}

The reduced $\Psi$-equation source still contains $\Theta''$, inherited from the
second scalar derivative in the Cotton stress. Using the scalar master equation
on shell,
\begin{equation}
\begin{aligned}
    &\frac{\Delta}{r^2}\frac{d}{dr}\!\left(\frac{\Delta}{r^2}\frac{d\Theta}{dr}\right)
+(\omega^2-V_\Theta)\Theta=0,\\
&V_\Theta=\frac{\Delta}{r^5}\Big[l(l+1)r+2M-\frac{2Q^2}{r}\Big],
\end{aligned}
\label{eq:appB-scalarEOM}
\end{equation}
to eliminate $\Theta''$, the source collapses to a purely algebraic coupling. The
coefficient of $\Theta'$ vanishes \emph{identically for all $Q$}, and one is left
with
\begin{equation}
\begin{aligned}
    &\Big[\tfrac{d^2}{dr_*^2}+\omega^2-V_{\rm RW}\Big]\Psi
= C_{\rm dCS}\,\Theta,\\
&C_{\rm dCS}=\frac{96\pi\,(Mr-Q^2)\,\Delta}{r^8}=\frac{96\pi\,(Mr-Q^2)\,f}{r^6}.
\end{aligned}
\label{eq:appB-result}
\end{equation}
Two checks confirm \eqref{eq:appB-result}. First, the $\Theta'$ coefficient was
verified to cancel symbolically at generic $Q$, so the coupling is genuinely
algebraic. Second, in the neutral limit,
\begin{equation}
C_{\rm dCS}\big|_{Q\to0}=\frac{96\pi M f}{r^5},
\label{eq:appB-Q0}
\end{equation}
\vspace{0.1cm}\\
reproducing the Schwarzschild dCS coupling of Molina \emph{et al.}~\cite{Molina2010}
[their Eq.~(2.8)]. The coupling is built from the combination $(Mr-Q^2)$ common to
all axial Cotton components (Appendix~\ref{app:cotton}), multiplied by the horizon
factor $\Delta$, and carries no explicit $l$-dependence.

\paragraph{Maxwell master equation.} The Maxwell equation $\nabla_\mu F^{\mu\varphi}=0$ involves $h_0$ and $h_0'$ [Eq.~\eqref{eq:CQcomponent}]. When $h_0$ is eliminated through the \emph{sourced} equations~\eqref{eq:appB-I}--\eqref{eq:appB-II}, the $C$-tensor source is carried into the Maxwell master equation. With $\Psi_e$ normalized as in Eqs.~\eqref{eq:H2coupled}--\eqref{eq:H1coupled}, this gives the term $\chi\,C_{\rm dCS}\,\Theta$ of Eq.~\eqref{eq:coupled-em}. The $\Theta'$ terms again cancel identically. The remaining ($t\varphi$) field equation, including its source, is then satisfied identically on shell; we verified this at generic $(M,Q,\omega,l)$ with exact rational arithmetic.

\subsection{Remark: why not the Cunningham--Price--Moncrief normalisation}
\label{app:red-cpm}

It is worth recording why the algebraic form \eqref{eq:appB-result} requires the
normalisation \eqref{eq:appB-Psi} rather than the Cunningham--Price--Moncrief
(CPM) master function used in the Martel--Poisson formalism. In the CPM
normalisation the master function itself carries a factor $2r/[(l-1)(l+2)]$, and
the corresponding source acquires both an explicit factor $1/[(l-1)(l+2)]$ and a
first-derivative ($\Theta'$) term. These are artefacts of the CPM normalisation,
not physical: they are removed by the $\Psi=i\Psi_{\rm RW}/\omega$ redefinition,
under which the $\omega$-dependence of the $r\varphi$ channel is absorbed and the
$1/[(l-1)(l+2)]$ factor cancels against the multipole weight of the reciprocal
coupling (Appendix~\ref{app:reciprocal}). The dipole $l=1$, for which
$(l-1)(l+2)=0$, is non-radiative and excluded from the master-variable
description, as usual.

\section{The reciprocal coupling and the electromagnetic channel}
\label{app:reciprocal}

\subsection{Linearized Pontryagin density}
\label{app:recip-direct}

The pseudoscalar obeys $\beta\,\Box\vartheta=-\tfrac{\alpha}{4}\pRR$. On the background $\pRR=0$, so the source is $\delta(\pRR)$ evaluated on the Regge--Wheeler-gauge perturbation $h_{t\varphi}=h_0(r)\,W\,e^{-i\omega t}$, $h_{r\varphi}=h_1(r)\,W\,e^{-i\omega t}$. Since $\pRR$ is linear in each Riemann factor and vanishes on the background, and since $\sqrt{-g}$ is unchanged at linear order for axial perturbations,
\begin{equation}
\delta(\pRR)=\varepsilon^{\rho\sigma\tau\eta}\,\bar R^{\nu}{}_{\mu\rho\sigma}\,\delta R^{\mu}{}_{\nu\tau\eta},
\end{equation}
where $\bar R$ is the background Riemann tensor. With the conventions of Sec.~\ref{sec:action_field_eq}, a direct computation gives
\begin{equation}
\delta(\pRR)=-\frac{24L\,(Mr-Q^2)}{r^7}\Big[r\,h_0'-2h_0+i\omega r\,h_1\Big]Y_l,
\label{eq:appC-dstarRR}
\end{equation}
exactly. All $\partial_\theta Y_l$ terms cancel on use of $W''=\cot\theta\,W'-LW$, as they must for a pseudoscalar source. In the neutral limit
\begin{equation}
\delta(\pRR)\big|_{Q=0}=-L\Big[\frac{24M}{r^5}h_0'-\frac{48M}{r^6}h_0+\frac{24\,i\omega M}{r^5}h_1\Big]Y_l .
\end{equation}

\subsection{Reduction}
\label{app:recip-reduce}

At the order required, $h_0$, $h_1$ and $\delta A_\varphi$ obey the general-relativistic Einstein--Maxwell equations.
\begin{itemize}
\item The $\theta\varphi$ component fixes $i\omega h_0=-(\Delta/r^5)\big[r\Delta h_1'+2(Mr-Q^2)h_1\big]$.
\item The $r\varphi$ component then gives Eq.~\eqref{eq:rphi-reduced}, where the $\Psi_e$ term is the electromagnetic stress $\delta T^{\rm EM}_{r\varphi}\propto Q\,\omega\,\delta A_\varphi$.
\item In the same variables the Einstein--Maxwell system takes the symmetric form~\eqref{eq:H2coupled}--\eqref{eq:H1coupled}. This fixes the normalization $\delta A_\varphi\propto(\mu/2)\Psi_e$ and reproduces $V_{\rm RW}$, $V_{\rm EM}$ and $C_Q$.
\end{itemize}
Substituting into Eq.~\eqref{eq:appC-dstarRR} gives Eq.~\eqref{eq:dstarRR-master}. With $\delta\vartheta=(\Theta/r)Y_l e^{-i\omega t}$,
\begin{equation}
\Box\,\delta\vartheta=\frac{1}{rf}\Big[\frac{d^2}{dr_*^2}+\omega^2-V_\Theta\Big]\Theta\,Y_l\,e^{-i\omega t},
\end{equation}
so
\begin{equation}
\Big[\frac{d^2}{dr_*^2}+\omega^2-V_\Theta\Big]\Theta=-\frac{\alpha}{4\beta}\frac{\Delta}{r}\,\delta(\pRR)\big|_{Y_l}
=C_{\Theta g}\big(\Psi+\chi\Psi_e\big),
\end{equation}
with $C_{\Theta g}$ as in Eq.~\eqref{eq:CThetag-result}.

\subsection{Checks}
\label{app:recip-calib}

\begin{itemize}
\item \emph{(i) Neutral limit.} $\chi\to0$ and $C_{\Theta g}\to(6M\alpha/\beta)\,[(l+2)!/(l-2)!]\,f/r^5$, which coincides with $V_{21}$ of Ref.~\cite{Molina2010} [their Eq.~(2.9)].
\item \emph{(ii) Reciprocity.} The gravity-side channels $C_{\rm dCS}$ and $C_{e\Theta}=\chi C_{\rm dCS}$ (Appendix~\ref{app:reduction}) and the scalar-side channels $C_{\Theta g}$ and $C_{\Theta e}=\chi C_{\Theta g}$ obey Eq.~\eqref{eq:CThetag-rel}. This is the structure required by the single bilinear $\tfrac{\alpha}{4}\int\sqrt{-g}\,\delta\vartheta\,\delta(\pRR)$.
\item \emph{(iii) Consistency.} The unused $t\varphi$ equation is satisfied identically on shell.
\item \emph{(iv) Dipole.} $C_{\Theta g}$ and $C_{\Theta e}$ vanish at $l=1$. For $l=1$ the gravitational sector is non-radiative and $\Psi_e$, normalized with $\mu$, must be replaced by the Maxwell potential itself; the coupled system~\eqref{eq:coupled} applies for $l\ge2$.
\item \emph{(v) Orientation.} Reversing the orientation flips the signs of all four dCS couplings together and leaves the frequencies unchanged.
\end{itemize}

\section{Local-WKB proxy for the fundamental-mode shift}
\label{app:wkb}

This appendix records a local-WKB estimate of the shift of the fundamental ($l=2$, $n=0$) mode. It combines leading-order WKB with a local (off-resonance) approximation for the pseudoscalar Green's function, and it omits the electromagnetic channel $\chi$. It is retained only as a qualitative proxy. The quantitative results are the coupled spectral solutions of Sec.~\ref{sec:wkb}.

\subsection{Method}
\label{app:wkb-method}

To leading WKB order the quasinormal frequency is fixed by the peak of the
potential,
\begin{equation}
\omega_0^2\simeq V_{\rm RW}(r_{\rm peak})
-i\Big(n+\tfrac12\Big)\sqrt{-2\,V_{\rm RW}''(r_{\rm peak})},
\label{eq:appD-wkbfreq}
\end{equation}
where the prime denotes $d/dr_*$ and $r_{\rm peak}$ is the maximum of
$V_{\rm RW}$. A small correction $\delta V_{\rm eff}$ to the effective potential
shifts the squared frequency by its value at the peak,
$\delta(\omega^2)\simeq\delta V_{\rm eff}(r_{\rm peak})$, since the unperturbed
mode is localised there and $V_{\rm RW}'(r_{\rm peak})=0$.

Integrating out the scalar in the coupled system
\eqref{eq:coupled}--\eqref{eq:allcouplings} generates the effective correction
$\delta V_{\rm eff}=C_{\rm dCS}\,G_\Theta\,C_{\Theta g}$, with
$G_\Theta=(\square+\omega_0^2-V_\Theta)^{-1}$ the scalar Green's function. In the
local (off-resonance) approximation---assumed here on the grounds that the gravitational and scalar peaks nearly coincide and the two channels are
non-degenerate---one replaces $G_\Theta\simeq(\omega_0^2-V_\Theta)^{-1}$
evaluated at the peak, giving the working formula
\begin{equation}
\delta(\omega^2)_{l,n}\simeq
\frac{C_{\rm dCS}\,C_{\Theta g}}{\omega_0^2-V_\Theta}\bigg|_{r_{\rm peak}},
\qquad
\delta\omega=\frac{\delta(\omega^2)}{2\omega_0}.
\label{eq:appD-shift}
\end{equation}

\subsection{The potential peak}
\label{app:wkb-peak}

For $l=2$ and $Q=0$ the Regge--Wheeler potential is
$V_{\rm RW}=f\big(6/r^2-6M/r^3\big)$ with $f=1-2M/r$; expanding,
$V_{\rm RW}=6/r^2-18M/r^3+12M^2/r^4$, so that $dV_{\rm RW}/dr=0$ gives
$2r^2-9Mr+8M^2=0$ and
\begin{equation}
r_{\rm peak}\big|_{Q=0}=\frac{9+\sqrt{17}}{4}\,M\simeq 3.2808\,M,
\label{eq:appD-peak}
\end{equation}
lying just outside the photon sphere $r_{\rm ph}=3M$. For $Q>0$ the peak of the
RN potential $V_{\rm RW}=(\Delta/r^5)[l(l+1)r-6M+4Q^2/r]$ moves inward with
charge; the values used below are obtained by locating the maximum numerically.

\subsection{Worked example: the neutral case}
\label{app:wkb-worked}

We display all intermediate quantities at $Q=0$ (units $M=1$, $\alpha=1$; the
physical shift is obtained on dividing by $\beta$). With
$r_{\rm peak}=3.28078$ and the continued-fraction--validated baseline
$\omega_0=0.373672-0.088962\,i$ (Sec.~\ref{sec:numval}), so that
$\omega_0^2=0.131716-0.066485\,i$, the ingredients of \eqref{eq:appD-shift} are
\begin{equation}
\begin{aligned}
V_\Theta(r_{\rm peak})&=0.239728,\\
\omega_0^2-V_\Theta(r_{\rm peak})&=-0.108012-0.066485\,i,\\
C_{\rm dCS}(r_{\rm peak})&=0.309767,\\
\beta\,C_{\Theta g}(r_{\rm peak})&=0.147903,
\end{aligned}
\label{eq:appD-ingredients}
\end{equation}
where $C_{\rm dCS}=96\pi(Mr-Q^2)\Delta/r^8$ and
$C_{\Theta g}=6(Mr-Q^2)(l-1)l(l+1)(l+2)\Delta/(\beta r^8)$ from
\eqref{eq:allcouplings}, and for $l=2$ the tensor weight is
$(l-1)l(l+1)(l+2)=24$. Assembling \eqref{eq:appD-shift},
\begin{equation}
\beta\,\delta\omega\big|_{Q=0}^{\rm WKB}\simeq-0.45+0.15\,i,
\label{eq:appD-Q0result}
\end{equation}
to be compared with the exact $-0.7733-0.3946\,i$ [Eq.~\eqref{eq:q0-result}]: the sign of the real part is correct, the sign of the imaginary part is not.

\subsection{Charge dependence}
\label{app:wkb-charge}

Repeating the construction across charge, using the validated baseline
$\omega_0(Q)$ of Sec.~\ref{sec:numval} at each point, gives the values in Table~\ref{tab:appD-charge}.
\begin{table}[htbp]
\centering
\begin{tabular}{cccc}
\toprule
$Q/M$ & $r_{\rm peak}/M$ & $\beta\,\delta\omega$ (local WKB) & $\beta\,\delta\omega$ (exact)\\
\midrule
$0.0$ & $3.28$ & $-0.45+0.15\,i$ & $-0.77-0.39\,i$\\
$0.2$ & $3.25$ & $-0.47+0.15\,i$ & $-0.76-0.39\,i$\\
$0.4$ & $3.16$ & $-0.53+0.16\,i$ & $-0.73-0.39\,i$\\
$0.6$ & $2.98$ & $-0.67+0.17\,i$ & $-0.72-0.38\,i$\\
$0.8$ & $2.70$ & $-1.0+0.21\,i$ & $-0.75-0.39\,i$\\
\bottomrule
\end{tabular}
\caption{Local-WKB proxy for the fundamental ($l=2$, $n=0$) shift, compared with the coupled spectral solution (units $M=\alpha=1$). The proxy fixes the order of magnitude and the sign of the real part only. It reverses the sign of $\mathrm{Im}\,\delta\omega$ and overestimates the growth with charge.}
\label{tab:appD-charge}
\end{table}
As $Q\to M$ the peak remains well outside the horizon $r_+\to M$, so the coupling
combination $(Mr-Q^2)|_{r_{\rm peak}}$ stays $O(1)$ and the fundamental
(photon-sphere) mode is \emph{not} suppressed; this is the family distinguished
from the near-horizon modes in Sec.~\ref{sec:extremal}.

\subsection{Accuracy}
\label{app:wkb-accuracy}

Two approximations enter Eq.~\eqref{eq:appD-shift}: leading-order WKB for the fundamental mode, and the local form of $G_\Theta$; in addition, the electromagnetic channel is omitted. The phase of $\omega_0^2-V_\Theta$ controls the sign of $\mathrm{Im}\,\delta\omega$, which the proxy gets wrong. The proxy also misses the non-monotonic charge dependence. Its robust outputs are only the order of magnitude and the sign of the real (frequency) shift.

\section{Numerical methods and convergence}
\label{app:numerics}

\begin{table*}[htbp]
\centering
\caption{Estimated numerical uncertainty of the quoted results (units $M=\alpha=1$).}
\label{tab:convergence}
\begin{tabular}{lll}
\hline\hline
Quantity & Parameters & Uncertainty\\
\hline
RN baseline $\omega_0$, $l=2$, $n=0$ & $Q/M\le0.8$ (Table~\ref{tab:numval}) & $<10^{-5}$ ($8\times10^{-9}$ at $Q/M\le0.1$)\\
$\beta\,\delta\omega$, $l=2$ & $Q/M\le0.98$ & \cell{grid/contour: $\le4\times10^{-5}$;\\ $1/\beta$ extrapolation: $\le4\times10^{-4}$}\\
Splitting at $Q_c$ & $l=2,3$, $\beta=10^3$--$10^{10}$ & five digits\\
ZDM exponent $p$ & $\sigma\to0$ & $2.00\pm0.01$\\
Eikonal ratios & $l\to\infty$ & $\le1.3\%$\\
$\beta\,\delta\omega$, $l=6$ & $Q/M\le0.98$ & $\le0.4\%$\\
ZDM coefficient $C(\sigma)$, $l=2$ & $\sigma=0.004$--$0.048$ & fifth digit\\
Extrapolated $C_0$ & $\sigma\to0$ & $\approx0.2\%$ ($481.7\pm0.6$)\\
\hline\hline
\end{tabular}
\end{table*}

\subsection{Exterior complex scaling}
\label{app:ecs}

In the normalization $\Psi=i\Psi_{\rm RW}/\omega$ the system~\eqref{eq:coupled} reads $\mathsf{H}\,\mathbf{v}=\omega^2\mathbf{v}$, with
\begin{equation}
\mathsf{H}=-\partial_{r_*}^2+\mathrm{diag}(V_{\rm RW},V_{\rm EM},V_\Theta)+\mathsf{C},
\end{equation}
where $\mathsf{C}$ contains the couplings and all entries are independent of $\omega$.
\begin{itemize}
\item \emph{Contour.} We continue $r_*=e^{i\theta}s$ with $s\in[-S,S]$. For $\theta>|\arg\omega|$ the quasinormal solutions $e^{\mp i\omega r_*}$ decay at both ends, so homogeneous Dirichlet conditions select the quasinormal spectrum. The continuum is rotated away from the modes.
\item \emph{Map $r(s)$.} The map is obtained by integrating $du/ds=e^{i\theta}(r-r_-)/r^2$ for $u=\ln(r-r_+)$ (DOP853, relative tolerance $10^{-13}$). This avoids logarithmic branch cuts.
\item \emph{Discretization.} Chebyshev collocation with $N+1$ points, and a dense eigensolve.
\item \emph{Shift extraction.} $\delta\omega(\beta)=\omega(\beta)-\omega(\infty)$ is computed on the same grid, so discretization errors largely cancel. The linear coefficient is extracted by Richardson extrapolation from $\beta=2\times10^4$ and $4\times10^4$. The difference between the two single-$\beta$ estimates, a conservative bound on the extrapolation error, is $\le4\times10^{-4}$ (absolute) for $l=2$ and $\le0.4\%$ (relative) for $l=6$.
\end{itemize}

\emph{Gates.}
\begin{itemize}
\item Schwarzschild $l=2$: $M\omega=0.3736716777-0.0889623110\,i$ (error $8\times10^{-9}$).
\item RN baseline: agrees with the charge series to $8\times10^{-9}$ at $Q/M=0.1$, and with Ref.~\cite{BertiKokkotas2005} at all charges (Table~\ref{tab:numval}).
\item The neutral shift~\eqref{eq:q0-result} agrees with the frequency-domain result of Ref.~\cite{Molina2010}.
\item The branch-reduced system~\eqref{eq:reduced} reproduces the full-system coefficients to $5\times10^{-7}$ ($l=2$) and $10^{-4}$ ($l=6$) at $\beta=2\times10^4$; the difference falls as $\beta^{-2}$ ($1.3\times10^{-4}$, $1.4\times10^{-6}$, $1.5\times10^{-8}$ at $\beta=2\times10^{4,5,6}$, $l=6$, $Q=0.2M$), i.e.\ it is $\mathcal{O}(\zeta^2)$.
\item Independent check: shooting integration of the full $\{\Psi,\Psi_e,\Theta\}$ system (Appendix~\ref{app:zdm} solver) gives $\beta\,\delta\omega$ within $1.4\times10^{-6}$ of the spectral values at $l=2$, $Q/M=0,0.6,0.9$, and $\omega_0$ within $8\times10^{-9}$.
\item Linearity in $\zeta$: at seven values $\beta=5\times10^3$--$3.2\times10^5$, $\beta\,\delta\omega=c_1+c_2/\beta+c_3/\beta^2$ fits with residual $\le6\times10^{-6}$, and $c_1$ equals the two-point Richardson value to $10^{-6}$.
\end{itemize}

\emph{Convergence.} We test the shift itself, not only the two frequencies. Varying $\theta\in[0.5,0.8]$, $S\in[45,120]$, $N\in[220,460]$ and the pivot of the rotation, $r_*=r_*^c+e^{i\theta}(s-r_*^c)$ with $r_*^c\in[-2,5]M$, changes $\beta\,\delta\omega$ by at most $4\times10^{-5}$ ($l=2$, $Q=0$), $3\times10^{-5}$ ($Q=0.6M$), $4\times10^{-9}$ ($Q=0.95M$), $2\times10^{-8}$ ($l=6$) and $7\times10^{-7}$ (electromagnetic-led, $Q=0.3M$). Smaller angles ($\theta=0.4$) require a larger box. Table~\ref{tab:convergence} summarizes the numerical uncertainty of each quoted quantity.

\subsection{Zero-damping modes}
\label{app:zdm}

Complex scaling cannot reach $\arg\omega\simeq-\pi/2$. Near the ZDMs, $\omega/\kappa_+$ is close to an integer, the horizon exponents $\pm i\omega/2\kappa_+$ differ by nearly an integer, and both Chebyshev and Leaver-recurrence methods degrade~\cite{Leaver1990RN}. We therefore use a shooting method in $r$:
\begin{itemize}
\item \emph{Horizon side.} The ingoing solutions are generated by a matrix Frobenius series to order 60. The Taylor coefficients of the regularized equation are obtained by FFT on the circle $|r-r_+|=(r_+-r_-)/2$, and the series is summed at $r-r_+=0.3(r_+-r_-)$. A near-integer exponent gap then produces only a large but finite coefficient, handled exactly.
\item \emph{Infinity side.} Outgoing solutions are started on the ray $r=8M+t\,e^{3\pi i/4}$, where $e^{i\omega r}$ decays for $\omega\simeq-i\gamma$, and are integrated inward, the stable direction.
\item \emph{Quasinormal condition.} The vanishing of the determinant of horizon and outgoing solutions at $r=r_++2(r_+-r_-)$.
\end{itemize}

\emph{Validation.}
\begin{itemize}
\item Photon-sphere modes are reproduced to $10^{-10}$, including the Schwarzschild $l=2$ $n=1$ overtone.
\item For the scaled ZDM residual $M(n+\tfrac12)^{-1}\sigma^{-2}\Delta\omega_I$ of Ref.~\cite{ZimmermanMark2016} (their Fig.~10) we obtain:
  \begin{itemize}
  \item $-3.02$ and $-1.35$ ($j=1$, $l=1$, $n=0,1$), against their $3.0$ and $1.35$;
  \item $-2.01$ and $-1.01$ ($j=2$, $l=2$), against their $2.0$ and $1.0$;
  \item $-1.69$ ($j=1$, $l=2$, $n=1$), against their $1.6$.
  \end{itemize}
\item Changing the matching radius or the Frobenius summation point alters the ZDM shift only in the fifth digit.
\end{itemize}

\emph{Code availability and AI use.} All codes and data used here are publicly available; see the Data availability statement~\cite{RNdCS_code}. The symbolic and numerical codes, including the plotting scripts for all figures, were written with the assistance of an AI tool, as described in the Acknowledgments.

\bibliography{references}
\end{document}